%% file: main.tex
\documentclass[lettersize,journal]{IEEEtran}
\usepackage{amsmath,amsfonts}
\usepackage{algorithmic}
\usepackage{algorithm}
\usepackage{array}
\usepackage[caption=false,font=normalsize,labelfont=sf,textfont=sf]{subfig}
\usepackage{textcomp}
\usepackage{stfloats}
\usepackage{url}
\usepackage{verbatim}
\usepackage{graphicx}
\usepackage{cite}
\usepackage{pifont}
\usepackage[table,dvipsnames, svgnames, x11names]{xcolor}
\usepackage{booktabs}
\usepackage{tcolorbox}
\tcbuselibrary{skins, breakable}
\usepackage{tikz}
\usepackage{collcell}

\begin{document}
\input{macros}

\title{Toward Understanding Operating System Defects}

\author{Hongyao Zuo, Jiali Li and Jiajun Jiang
\thanks{Hongyao Zuo and Jiali Li are joint first authors; Jiajun Jiang is the corresponding author.}
\thanks{Hongyao Zuo, Jiali Li and Jiajun Jiang are with School of Computer Software, College of Intelligence and Computing, Tianjin University, Tianjin 300350, China. (E-mail: zuohongyao@tju.edu.cn; jlfi77@tju.edu.cn; jiangjiajun@tju.edu.cn).}}

\markboth{Journal of \LaTeX\ Class Files,~Vol.~1, No.~1, August~2026}%
{Shell \MakeLowercase{\textit{et al.}}: A Sample Article Using IEEEtran.cls for IEEE Journals}


\maketitle

\begin{abstract}

Operating systems (OS) serve as the foundation for all other software systems, and thus defects in OSes can lead to severe conquences, such as system crashes and data corruption, affecting billions of users. This broad impact underscores the necessity and importance of ensuring OS quality. Understanding the characteristics of OS defects is a fundamental step in this quality assurance task, as it facilitates the design of effective defect detection and debugging approaches. In this work, we conduct a large-scale study of 1,500 defects from three distinct and representative operating systems (Android, Linux, and HarmonyOS) spanning both mobile and desktop environments. To the best of our knowledge, this is the largest study of its kind in this domain. By analyzing the distribution of OS defects across multiple classification dimensions, including the OS layer where defects occur, the functions they affect, how they are triggered, their severity, and the code elements involved in their repair, as well as performing joint analysis between dimensions and cross-OS similarity comparisons, we summarize several major findings that contribute to a comprehensive understanding of OS defects across systems. Based on these findings, we provide a series of actionable implications for better OS defect detection and debugging, offering guidelines for future research aimed at improving OS quality assurance.

\end{abstract}


\input{intro}
\input{related}
\input{study}
\input{analysis}

\input{implication}

\input{discussion}
\input{conclusion}




\bibliographystyle{IEEEtran}
\bibliography{references}

\end{document}

%% file: macros.tex
\newcommand{\jiajun}[1]{\textcolor{orange}{\ding{46}[Jiajun:#1]}}

\newcommand{\codeIn}[1]{\texttt{#1}}

\newcommand{\modify}[1]{\textcolor{purple}{[modify:#1]}}

\newtcolorbox[auto counter]{finding}[1][]{%
    enhanced,
    breakable,
    colback=gray!8,          
    colframe=gray!40,        
    boxrule=0.6pt,           
    arc=4pt,                 
    outer arc=4pt,
    before skip=1em,
    after skip=1em,
    left=6pt, right=6pt, top=6pt, bottom=6pt,
    title={\textbf{Finding \thetcbcounter}:},
    fonttitle=\bfseries,
    coltitle=black,
    attach title to upper={\ },
    #1
}

%% file: intro.tex
\section{Introduction}
\label{sec:intro}

\IEEEPARstart{O}perating systems (OS) serve as the foundational infrastructure for modern computing, underpinning everything from high-performance servers to ubiquitous mobile devices. As the complexity of these systems continues to grow, ensuring their reliability and security has become a paramount concern~\cite{1631939, 8662611}. A single defect in an operating system can have catastrophic consequences, ranging from data corruption and system crashes to large-scale security breaches that compromise the privacy and safety of billions of users~\cite{3277203.3277276, 10.1145/502034.502042}. Therefore, the systematic analysis and understanding of OS defects are critical for guiding quality assurance efforts and informing the development of more resilient system architectures.

Over the decades, the study of software defects in operating systems has evolved from analyzing monolithic kernels to understanding the intricacies of modern, heterogeneous OS landscapes. Early foundational work, such as the Orthogonal Defect Classification (ODC) framework invented by IBM Research, provided a seminal methodology for categorizing and analyzing defects in large-scale software systems~\cite{177364}. This framework has been widely adopted for in-process quality measurement and has influenced much of the subsequent research in this domain.

With the proliferation of mobile computing, research attention has increasingly shifted toward mobile operating systems. The Android ecosystem, due to its open-source nature and extensive deployment, has been the subject of numerous empirical studies, many of which have focused on security vulnerabilities~\cite{10.1145/2508859.2516728, 10.1145/1999995.2000018, 6234407}. However, these studies often focus on a single OS or a specific type of defect. In contrast, a comprehensive understanding of defects across different OS families, such as mobile, desktop, and specialized embedded systems, remains limited. Comparative analyses that do exist are often confined to performance benchmarking or high-level security evaluations, lacking a systematic, multi-dimensional exploration of defect patterns~\cite{10.1145/2745802.2745808, tan2014bug}. Furthermore, methodological frameworks for consistent and comparative defect analysis across fundamentally different operating system architectures are not well-established.

Given this research gap, we believe that a comprehensive, multi-faceted empirical study is needed to understand the characteristics of OS defects. Such a study is a fundamental prerequisite for designing effective defect detection, localization, and repair techniques tailored to the unique properties of modern operating systems. Without a deep understanding of the root causes, symptoms, and distribution of these defects, testing and debugging efforts will remain ad hoc and potentially ineffective, mirroring the challenges observed in other complex software domains like deep learning frameworks~\cite{10.1145/3587155}.

To address this issue, we conduct a large-scale empirical study on 1,500 real-world defects collected from three distinct operating systems: Android, Linux and HarmonyOS, covering different computing domains of both mobile computing and desktop  devices, and involving diverse mature levels of both long-standing well maintained and new emerging systems. Specifically, we curated these data from open-source platforms and conducted manual labeling to establish a unified classification framework for the identified defects. Building on this foundation, our comprehensive cross-system defect analysis pursues the following research questions, intended to offer a holistic understanding of OS defects:

\begin{description}
\item[RQ1:] How are defects distributed across different classification dimensions for each operating system?

\item[RQ2:] How do defect distributions correlate between different dimensions?

\item[RQ3:] To what extent are defect distributions similar across operating systems?
\end{description}

By addressing these research questions, our study makes the following major contributions:

\begin{itemize}
\item \textbf{Large-scale, Multi-OS Empirical Study:} We present a comprehensive study of 1,500 defects from three diverse operating systems: Android, Linux, and HarmonyOS, making it, to the best of our knowledge, one of the largest and most diverse studies of its kind, and summarize a set of findings to guide future research.
\item \textbf{Multi-Dimensional Defect Taxonomy:} We develop a multi-dimensional classification framework for OS defects, identifying their root causes (the triggers of the defects) and symptoms (the functions they affected and how to repair them), and analyze their distribution across systems.
\item \textbf{Actionable Implications for OS Quality Assurance:} Based on our findings, we propose a set of implications for improving the detection, debugging, and testing of operating system defects.
\item \textbf{A Foundation for Future Research:} Our findings and publicly released dataset aim to provide a solid foundation for future research on OS reliability and security, enabling the development of more targeted and effective quality assurance tools. All our experimental data are availabe at \textbf{https://zuohongyao.github.io}.
\end{itemize}

The remainder of this paper is organized as follows. Section~\ref{sec:related} discusses the related work and summarizes the necessity of this study. Section~\ref{sec:study} details our research methodology, including data collection and classiffication processes. Section~\ref{sec:result} presents a cross-system analysis of the results, based on which Section~\ref{sec:implication} summarizes the implications for developers and future research. Finally, Section~\ref{sec:threats} discusses threats to validity, while Section~\ref{sec:conclude} concludes the paper.

%% file: related.tex
\section{Related Work}
\label{sec:related}

Operating systems are the foundation of all other software systems, making it crucial to understand the characteristics of OS bugs. Over the past decades, extensive research has been conducted on this topic. Chou et al.~\cite{10.1145/502034.502042} analyzed Linux kernel snapshots and found that driver error rates are three to seven times higher than other kernel components. Rastogi et al.~\cite{ruohonen2024fastfixesfaultydrivers} confirmed that drivers remain the most regression-prone subsystem in the Linux kernel. Ren et al.~\cite{8377690} employed topic modeling on 240k Ubuntu bug reports and identified three general bug types. Yoshimura et al.~\cite{Yoshimura2013} analyzed error propagation in Linux using fault injection techniques. Xiao et al.~\cite{8731682} examined 5,741 Linux kernel bug reports from an evolutionary perspective and summarized 22 findings. Mu et al.~\cite{Mu2022AnIA} analyzed the causes and costs of duplicated kernel bug reports. Zhou et al.~\cite{zhou2026tamingcomplexitydemystifyingsoftware} proposed an LLM-based fault localization benchmark for the Linux kernel. Liu et al.~\cite{liu2025characteristicsrootcausesdetection} conducted the first study on incomplete security patches in the Linux kernel. O'Sullivan et al.~\cite{11291433} identified RCU synchronization issues in Linux drivers. Hu et al.~\cite{DBLP:conf/ndss/HuDLML26} systematically analyzed Linux kernel hardening techniques, revealing 23 unprotected attack vectors. Chen et al.~\cite{10.1145/3689031.3717487} introduce Seal, a security-patch-driven specification inference framework that identifies 167 new Linux kernel bugs with 71.9\% precision. He et al.~\cite{He2027HACMony} focused on audio-stream conflict issues in HarmonyOS and proposed an automated detection approach.

On the other hand, the proliferation of mobile computing has driven a surge in research on Android and iOS security. In a large-scale study of over 10k apps, Schmidt et al.~\cite{10.1145/3719027.3765033} found that iOS apps are more likely to expose sensitive data than Android apps. Wu et al.~\cite{10.1145/2508859.2516728} examined how vendor customizations affect Android security, while Steinböck et al.~\cite{11129415} surveyed mobile hardening techniques and reported that iOS implements only half as many hardening measures as Android.

Despite the extensive body of research, several limitations remain: (1) Most studies focus on fault injection or vulnerability patterns, lacking comprehensive defect characteristic analysis; (2) The majority target individual systems without cross-system comparison; (3) They concentrate on mature systems while paying insufficient attention to emerging architectures such as HarmonyOS; (4) and the available data for joint multi-system defect analysis is limited. To address these gaps, this study conducts the first comprehensive comparison across HarmonyOS, Android, and Linux, providing a multi-dimensional classification framework covering OS layers, affected functions, trigger scenarios, and quantitative metrics to systematically characterize defect distribution and latent defect patterns across different operating systems.

%% file: study.tex
\section{Methodology}
\label{sec:study}

To achieve a rigorous and systematic analysis of the defect patterns across different operating systems, this section introduces details of the data collection, and the defect classification and labelling process in detail, which ensures the validity and reliability of our empirical study.


\subsection{Data Collection}
\label{subsec:data}

To ensure a representative and analytically meaningful comparison of defect distribution across operating systems, the following criteria will guide the selection of candidate operating systems. Each criterion (except the first one) addresses a key dimension of variability that may influence defect patterns.

\begin{itemize}
	\item \textbf{Availability:} This criterion requires that the candidates must have a publicly accessible, well‑maintained bug database (e.g., Bugzilla, GitHub Issues, Jira) with consistent categorization to allow reliable extraction of defect distribution data. In addition, a sufficiently large number of bug reports should be available included to enable statistical analysis.

	\item \textbf{Mature Level:} This criterion requires the selected candidates include both long‑standing, mature OSes and emerging OSes for assessing whether the age and evolution of an OS are associated with systematic differences in defect distribution (e.g., defect types, severity, or frequency). The reason is that the long‑standing, mature OSes (e.g., Windows and Linux) have undergone extensive testing, real‑world deployment, and iterative bug fixing over many years. Their defect distributions may be dominated by corner cases, legacy code issues, or complex interactions. In contrast, emerging OSes (e.g., Redox OS and recent versions of HarmonyOS) are less mature, with smaller codebases and shorter field exposure. They may exhibit different defect patterns, such as a higher density of memory safety defects or rapid evolution of defect types. Comparing these two groups helps determine whether OS maturity is a significant factor in defect distribution. For instance, whether mature systems show a shift toward logic errors or concurrency defects, while emerging systems show more implementation or resource defects.
	
	\item \textbf{Computing Domain:} This selection criterion requires the selected candidates include both desktop and mobile operating systems for investigating whether the target computing environment (desktop vs. mobile) influences defect distribution, given differences in hardware, power constraints, user interaction models, and software ecosystems.
	The reason is that desktop OSes (e.g., Windows, macOS, mainstream Linux distributions) typically run on x86/x64 hardware with abundant power and cooling. They support complex, long‑running processes, extensive peripheral drivers, and legacy APIs. Defects may relate to resource management, peripheral I/O, or multi‑window UI. In contrast, mobile OSes (e.g., Android and iOS) operate on ARM hardware with strict power and thermal budgets. They emphasize touch interfaces, app sandboxing, and background process restrictions. Defect distributions may differ, e.g., higher prevalence of power‑management defects, inter‑process communication issues, or security defects due to app ecosystem constraints. Including both domains enables cross‑domain comparison and helps identify whether defect distribution is primarily domain‑specific or exhibits OS‑independent patterns.
	
	\item \textbf{Implementation Language:} This selection criterion requires the candidates involve operating systems written in different programming languages for evaluating whether the languages used for OS kernel and core components affect defect distribution, particularly with respect to memory safety, concurrency, and abstraction overhead. The reasons is that operating systems are implemented in various system programming languages. For example, C/C++ includes frequent memory management and pointer arithmetic, which can lead to buffer overflows, use‑after‑free, and other memory corruption defects. While Java is not typically used for kernel development, it plays a significant role in the user‑space of certain mobile operating systems (e.g., Android’s framework and application layer). In such contexts, defects may relate to garbage collection overhead, memory leaks due to lingering object references, or concurrency issues arising from the Java Memory Model. Although Java’s runtime safety reduces classic memory corruption defects.
	By including OSes written in different languages (or significant subsystems and components in different languages), the study can test whether language choice correlates with the prevalence of certain defect categories. This is particularly important for informing language adoption in system development.
	
\end{itemize}

This structured selection ensures both internal validity (clear comparisons across defined dimensions) and external validity (generalizability to real‑world OS development). According to these selection criteria, we finally selected three representative operating systems at the target in this study, including \textbf{Android}, \textbf{Linux}, and \textbf{HarmonyOS}, which cover all the aspects mentioned above. Specifically, the data collection process was tailored to the specific characteristics of each operating system's public defect repositories, ensuring comprehensive coverage and consistency across platforms. Before formally commencing data collection, we carefully studied the defect publication formats across the three operating systems. Defects published on Launchpad (for Linux) are not necessarily guaranteed to have been fully resolved, and patches are presented in the form of patch links within the bug reports. In contrast, the defects published for HarmonyOS and Android are all confirmed to be fully resolved. 
Ultimately, we extracted the common attributes shared across the three defect formats, namely bug ID, severity level, and patch, as the principles guiding our data collection. Other attributes that were not included in the final retained set were nonetheless crawled whenever present in the bug reports to support subsequent analysis. For each system, we adopted a systematic approach to gather defect instances. The detailed processes are presented as follows.

\begin{itemize}
    \item For \textbf{Android}, we developed a dedicated web crawler to collect defect data from publicly available Android Security Bulletins\footnote{https://source.android.google.cn/docs/security/bulletin}. These bulletins served as the primary data source, providing authoritative records of security-related defects. The crawler, implemented in Python, was designed to systematically extract detailed information for each vulnerability, including affected modules, CVE identifiers, defect IDs, severity ratings, patch information, disclosure dates, and associcated descriptions.

	\item For \textbf{Linux}, which hosts projects primarily for Ubuntu and related Linux distributions, we utilized the Advanced Search functionality provided by the Lanchpad platform\footnote{https://bugs.launchpad.net/}. We configured the search with the following parameters: (1) \textit{Status} set to ``Fix Released'' to include only resolved defects, and (2) \textit{Sort Order} set to ``Newest First'' to prioritize recent issues. For each defect matching the search criteria, we extracted the bug ID, descriptions and discussions, severity level, and patch links. 
	
	\item For \textbf{HarmonyOS}, we identified the official repositories maintained by Huawei and the open-source community on Gitee as the primary data sources\footnote{https://gitee.com/openharmony/security/tree/master/zh/security-disclosure}. We systematically mined these repositories to collect bug reports and security vulnerability disclosures. For each publicly disclosed defect, we extracted comprehensive information, including the defect ID, associated bug report, severity level, patch details, commit timestamps, and so on.

\end{itemize}

\noindent
\begin{table}[t]
\centering
\caption{Summary of collected bug data across three operating systems}
\label{tab:data_summary}
\resizebox{0.48\textwidth}{!}{%
\begin{tabular}{lccc}
\toprule
\textbf{OS} & \textbf{Time Range} & \textbf{Initial Bug Reports} & \textbf{Final Bugs} \\
\midrule
Linux & 2023.04 -- 2025.06 & 32,249 & 500 \\
HarmonyOS & 2023.05 -- 2025.06 & 712 & 500 \\
Android & 2023.05 -- 2025.12 & 1,750 & 500 \\
\bottomrule
\end{tabular}
}
\end{table}

After obtaining the initial bug reports (as shown in Table~\ref{tab:data_summary}), we implemented several measures to ensure data quality and consistency across the final dataset. We first performed deduplication by using defect IDs to identify and remove duplicate entries from all data sources, and all reports without open-source patch diffs were excluded from further analysis. 
After this initial filtering,  we further removed defects with code changes exceeding 500 lines, as they usually involve large refactoring rather than localized bug fixes.
For HarmonyOS and Linux, a special case was observed where multiple related defects were sometimes submitted together as a single combined report. In such cases, we analyzed each defect separately rather than treating the combined report as one data point. After applying all these filtering steps, we selected the latest 500 valid defect instances from each operating system for our study in the subsequent analysis. The details of the collected data are presented in Table~\ref{tab:data_summary}, in which we report the time range involved of the collected bug reports and the number of bugs finally analyzed for each OS.

\subsection{Classification and Labeling Process}
\label{subsec:process}

\begin{figure*}[!tbp]
  \centering
  \includegraphics[width=\textwidth]{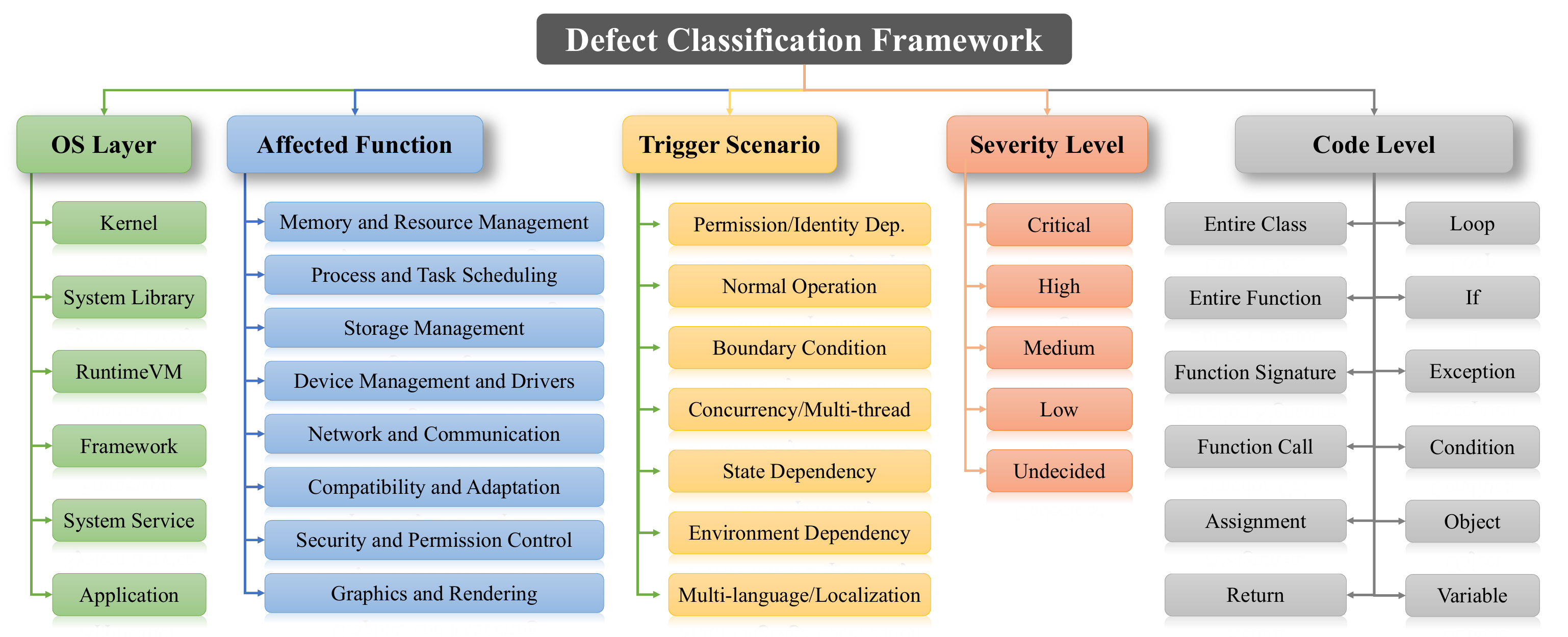}
  \caption{Schematic diagram of the defect classification framework constructed in our study.}
  \label{fig:overall_perf}
\end{figure*}

To ensure systematic and consistent defect classification across all three operating systems, we developed a comprehensive multi-dimensional classification framework for the final dataset, designed to capture multiple dimensions of defect characteristics for cross-platform comparison.

Prior to conducting the full classification, we executed a preliminary analysis phase in which we randomly selected 10 defect instances from each operating system (30 in total) to develop and refine the classification methodology. For each sampled defect, two annotators performed manual classification, meticulously examining each defect by carefully reviewing available bug reports and patch details to ensure classification accuracy. In this process, they were dedicated to identifying classification criteria and defining classfication labels by following the standard open coding procedure~\cite{Corbin1990Grounded}. To ensure quality and reliability, the initial phase involved both annotators independently classifying the same dataset, and any inconsistencies were resolved through thorough discussions by involving a third senior researcher to reach a consensus on the final classification. Based on this preliminary analysis, we refined the classification categories and established explicit criteria for each dimension, thereby finalizing an initial version of the multi-dimensional classification framework.

After the initial framework was established, we adopted an iterative annotation strategy to further refine the classification scheme. In each iteration, two annotators with at least three development experience independently labeled 50 defect instances following the current framework and subsequently analyzed the annotation results. The framework was continuously improved based on practical labeling experiences. During this process, we calculated inter-rater reliability and agreement statistics for categorical coding using the Cohen's Kappa coefficent~\cite{5584447} after labeling every 50 defects. As the classification process progressed, the overall Kappa coefficient gradually increased from an initial moderate level (0.45$\sim$0.60) to a substantial or near-perfect level (above 0.80) after two iterations of labeling, indicating improved precision and consistency of the classification framework over time. As a result, the two anotators labeled the remaining defects independently in the subsequent process and discussed inconsistencies with the third researcher for reaching a consensus. In particular, the Cohen's Kappa coefficent ranges from 0.8 to 1.0 in the subsequent labeling process, which ensures a consistent and reliable classification framework, yielding a high-quality, multi-dimensional dataset capable of supporting rigorous comparative analysis of defect patterns across the three
operating systems. This dataset serves as the foundation for the empirical findings presented in subsequent sections of this paper.

\subsection{Defect Classification Framework}
\label{sec:framework}

According to the previous explained labeling process, we constructed our final classification framework, which is presented in Figure~\ref{fig:overall_perf}. Specifically, each defect instance is classified from five distinct dimensions, including the OS layers that the defect occurred (\textit{OS Layer}), the primary functions affected by the defect (\textit{Affected Function}), the scenario where the defect could be triggered (\textit{Trigger Scenario}), and the severity of the defect labeled by the developers or maintainers (\textit{Severity}) and the code changed for repairing the defect (\textit{Code Level}). Except for the Code Level, the categories under each dimension are mutually exclusive. That is, every defect is assigned to exactly one category for any given dimension. In the following, we introduce each of the demensions and their associated categories in details.

\subsubsection{\underline{OS Layer}}

This dimension classifies defects according to the system layer where the root cause resides. The distinction between different operating system layers is typically determined by the location of the modified code. According to the architecture of different operating systems, the corresponding OS layer can be determined by the directory path of the modified code. This dimension comprises the following six categories.


\textbf{Kernel Layer:} Defects involving core kernel functionality, kernel-mode drivers, and the hardware abstraction layer (HAL), representing logic errors in core modules that directly control hardware and provide fundamental low-level system capabilities. The root cause lies in low-level hardware interaction implementations, including both kernel-mode and non-kernel-mode drivers. \textit{Characteristics:} Code typically located in the \codeIn{kernel/} directory (e.g., \codeIn{kernel/linux/}) , kernel modules or HAL libraries (e.g., \codeIn{hal/}). Functionality usually involves process creation and destruction, memory allocation, and system call implementations. \textit{Manifestations:} System crashes, deadlocks, and resource leaks (e.g., \codeIn{AndroidRuntime: FATAL EXCEPTION (kernel)}, Ubuntu: \codeIn{Kernel panic} and HarmonyOS: \codeIn{kernel restart}). 
These defects are typically triggered in the kernel mode, such as exposed through system calls.

\textbf{System Library Layer (SystemLib):} Defects located in fundamental libraries above the kernel, providing standardized interfaces for applications and frameworks, delivering basic functionality to upper layers such as mathematical operations and graphics rendering. \textit{Characteristics:} Code typically located in \codeIn{system/} directory library modules. Callers may be from the framework layer or system service layer, but do not involve application components. Functionality includes general utility classes (e.g., string processing and file I/O) or specialized domain libraries (e.g., graphics rendering and media codecs). \textit{Manifestations:} Library functional errors (e.g., missing parameter validation and incorrect return values) lead to failures whenever the library function is invoked, regardless of the caller or the programming language used by the caller (e.g., a buffer overflow in function `\codeIn{strcpy}' from Ubuntu \codeIn{glibc} causes all programs that invoke `\codeIn{strcpy}' to crash).

\textbf{Runtime/Virtual Machine Layer (RuntimeVM):} Defects involving core mechanisms of application runtime environments, directly related to bytecode execution, memory management, and virtual machine scheduling. \textit{Characteristics:} Code located in \codeIn{art/} directory (ART VM), \codeIn{dalvik/} directory (legacy version), or \codeIn{runtime/} related modules. It primarily affects applications dependent on that runtime (e.g., Android ART GC vulnerabilities causing all Java applications to crash frequently and HarmonyOS ArkVM parsing errors causing ArkTS application crashes). The involoved functionalities usually involve garbage collection algorithms, JIT/AOT compilation optimization, bytecode interpreters, and thread synchronization mechanisms. \textit{Manifestations:} It is related to the runtime environment rather than application logic (different applications will exhibit identical issues).

\textbf{Framework Layer:} Defects located in the core implementations of frameworks, providing modular functionality encapsulation, affecting application component lifecycle, system interface invocation, and other fundamental logic. \textit{Characteristics:} Code located in \codeIn{frameworks/base/} directory under non-system service layer directories. It primarily affects applications using that framework (e.g., Android \codeIn{RecyclerView} refresh logic errors cause application latency and HarmonyOS ArkUI layout engine defects cause interface disarray), usually involving functionalities about application component management (e.g., activity startup flow and service binding mechanisms), resource management (e.g., layout parsing and theme adaptation). \textit{Manifestations:} Framework API design defects (e.g., interface call sequence errors cause crashes), event handling logic errors (e.g., click event delivery anomalies).

\textbf{System Service Layer:} Defects located in independently running system service processes, typically involving cross-application core functionality scheduling. \textit{Characteristics:} Code located in \codeIn{frameworks/base/services/} directory, which involves system-level resource scheduling. \textit{Manifestations:} Multi-application interaction anomalies (e.g., system latency while switching applications caused by scheduling errors in \codeIn{ActivityManagerService}). This mainly affects the user applications and service restart can temporarily restore the functionality.

\textbf{Application Layer:} Defects located in those independent application code that are included in the system by default (such as camera in Android), affecting only single application functionality, while not involving system-level frameworks or services. \textit{Characteristics:} Code located in application APK source code (e.g., packages/apps/ directory under Browser, Settings). Functionality involves application-specific logic (e.g., button click event handling, data display errors), not involving cross-application or system-level interface low-level implementations. \textit{Manifestations:} Unrelated to underlying layers (after verifying system library and framework normality, can be classified). Problems only reproduce in specific applications (e.g., browser unable to load webpages due to application-internal network request logic errors, Ubuntu gedit unable to save files while other editors function normally; HarmonyOS camera application crashes while third-party camera applications function normally).

\subsubsection{\underline{Affected Function}}

This dimension classifies defects according to the primary functions affected by each defect, comprising the following eight categories:

\textbf{Memory and Resource Management (Mem/Res Mng):} Defects involving system/application memory allocation, deallocation, leaks, and hardware resource scheduling such as CPU/GPU, covering lifecycle management issues for memory and various hardware resources. \textit{Core modules involved:} libc (memory allocation low-level library), art (VM memory management), and related modules. \textit{Typical manifestations:} Memory anomalies including memory leaks, memory out-of-bounds access; resource anomalies including file handle leaks causing resource exhaustion, and so on.

\textbf{Process and Task Scheduling (Pro/Tas Sch):} Defects involving process creation, destruction, priority scheduling, task stack (TaskStack) and foreground/background transitions, usually affecting system task execution order and CPU time allocation. \textit{Core modules involved:} \codeIn{TaskRecord} (managing task stacks), \codeIn{WindowManagerService} (managing GUI tasks), and related modules. \textit{Typical manifestations:} Process anomalies including process unresponsiveness, abnormal restarts; task anomalies including background task exceptional termination, task stack switching failures, and so on.

\textbf{Storage Management (Sto Mng):} Defects involving both internal storage (e.g., memories and file systems) and external storage (e.g., flash cards), affecting data access and storage device management. \textit{Core modules involved:} \codeIn{kernel/fs}(file system drivers), \codeIn{FileProvider} (file access control), and related modules. \textit{Typical manifestations:} Storage anomalies including flash card mounting failures, file access permission errors (e.g., unable to read album images), incorrect storage capacity calculation (e.g.,  display inaccurate available space); data write failures including failures on synchronizing data to disks through \codeIn{SharedPreferences}), and file system corruptions causing data inaccessibility.

\textbf{Device Management and Drivers (Dev/Drv):} Defects involving hardware device (e.g., cameras, sensors, Bluetooth, etc.) driver interactions and HAL layer (Hardware Abstraction Layer) adaptation, affecting hardware availability and data accuracy. \textit{Core modules involved:} \codeIn{frameworks/opt/bluetooth} (Bluetooth drivers), \codeIn{hardware/camera} (camera drivers), and related modules. \textit{Typical manifestations:} Device unavailability including camera startup failures, Bluetooth connection failures, sensor non-responsiveness; device data anomalies including inaccurate sensor values, Bluetooth audio transmission lag, and so on.

\textbf{Network and Communication (Net/Com):} Defects involving network protocols (TCP/IP, HTTP), network connection management (WiFi, cellular networks), and data transmissions, affecting network connection stability and data interaction accuracy. These are system-level defects caused by implementation flaws in protocol handling, state machine mismanagement, or improper error recovery, rather than routine environmental issues (e.g., weak signal or server downtime). \textit{Core modules involved:} \codeIn{WifiManager} (WiFi management), \codeIn{libcore/net} (protocol stack), and related modules. \textit{Typical manifestations:} Network connection anomalies including WiFi scan failures, cellular network registration failures, network switching failures; data transmission anomalies including data packet loss, HTTP request timeouts, and so on.

\textbf{Compatibility and Adaptation (Com/Adp):} Defects that manifest as functional incompatibility when the systems or included applications operate across different hardware platforms, API environments, or different system versions, affecting cross-version and cross-device functionality consistency. \textit{Typical manifestations:} Version compatibility issues including old applications crash in latest-released systems (e.g., calling deprecated APIs without adaptation), new system features unavailable on old devices; hardware compatibility issues including application behavioral inconsistencies across devices of different architectures (e.g., graphics rendering errors on x86, but not on ARM).

\textbf{Security and Permission Control (Sec/Perm):} Defects involving permission validation, data encryption, and malicious behavior protection (e.g., SELinux, sandboxing), affecting system data security and access compliance. \textit{Core modules involved:} \codeIn{Keystore} (key storage) and related modules. \textit{Typical manifestations:} Permission boundary issues including unauthorized applications accessing sensitive data (e.g., contacts and locations), permission validation bypass (e.g., invoking permission-requiring services through implicit intents); security vulnerabilities including encryption algorithm key leaks, SELinux policy configuration errors causing malicious application privilege escalations, and so on.

\textbf{Graphics and Rendering (Gra/Ren):} Defects involving UI rendering, animation effects, and so on, affecting interface display and interaction fluency. \textit{Core modules involved:} \codeIn{libhwui} (UI rendering engine), \codeIn{hardware/gpu} (OpenGL/Vulkan engine), and related modules. \textit{Typical manifestations:} Display anomalies including graphical element misalignment (e.g., button overlaps); rendering performance issues including frame lag and frame drops, and so on.

\subsubsection{\underline{Trigger Scenario}}

Defects are classified according to the manners and scenarios in which defects are triggered. This dimension comprises the following seven categories:

\textbf{Permission/Identity Dependency Trigger (Per/Ide Dep):} Defects triggered only under specific user permissions or identity roles. The same operation executed under different roles exhibits different behaviors (e.g., administrators can trigger anomalies while regular users experience no issues). For example, in \codeIn{WindowManagerService}, the permission check was misplaced in the internal \codeIn{doDump()} method rather than the public \codeIn{dump()} entry point, enabling unauthorized callers to bypass it by supplying the `\codeIn{--high-priority}' flag when invoking \codeIn{PriorityDump.dump()}.

\textbf{Regular Operation Trigger (Reg Ope):} Defects triggered during regular system functionality usage (e.g., clicking buttons, invoking APIs), requiring no special environment, permission, or boundary value. It is reproducible through single or combined operational flows.
For example, the remote speech recognition service incorrectly included the \codeIn{BIND\_ALLOW\_BACKGROUND\_ACTIVITY\_STARTS} flag when binding to third-party speech apps, allowing malicious apps to bypass system restrictions and illegally launch activities in the background. It is a routine operation triggered security flaw as whenever users use related functions it may occur.

\textbf{Boundary Condition Trigger (Bor Con):} Defects triggered under extreme inputs (null values, excessively large numbers, illegal formats), resource exhaustion (low memory, file descriptor exhaustion, thread pool exhaustion), or exceptional flows (network interruption, forced process termination, excessive recursion) where the root cause lies in the defective handling of boundary conditions within the source code. This type of defects usually expose issues such as missing input validation, insufficient resource management, or inadequate exception handling. For example, when the query interface of \codeIn{AppOpsService} is invoked, the system service does not limit the number of returned entries. A malicious app deliberately creates a large number of attributed permission entries, causing the response data to exceed the Binder transaction size limit, thereby triggering an exception and resulting in a denial of service error.

\textbf{Concurrency/Multi-thread Trigger (Con/Mul Thr):} Defects triggered by multi-thread race conditions, lock mechanism errors, or asynchronous task processing anomalies. They have to be reproduced under multi-threaded concurrency or asynchronous tasks rather than a single-threaded environment. An example from Android shows that a race condition arose when \codeIn{finishTransition()} updated the \codeIn{InputSink} state using a separate transaction instead of the pending one, causing the input sink region to be misaligned with the actual window position whenever the two transactions interleaved in execution.

\textbf{State Dependency Trigger (Sta Dep):} Defects triggered by system state changes, including time progression (e.g., timeout of scheduled tasks); configuration/parameter updates (e.g., \codeIn{config.xml} modifications); runtime state transitions (e.g., background-to-foreground transitions and screen rotations). Taking an Android defect as an example, when an \codeIn{Activity} transition from the \textit{resumed} state (fully visible and interactive) to the \textit{paused} state (partially obscured), the \codeIn{onPause()} method is called. In this case, due to the missing security flag in \codeIn{onPause()} that was meant to prevent tapjacking, where a malicious app overlays a transparent, fake interface over a legitimate confirmation button to trick users into tapping it without realizing, an attacker could then overlay a decoy button and fool the user into granting unintended permissions during the state transition process.


\textbf{Environment Dependency Trigger (Env Dep):} Defects triggered by external environmental factors, including network environments (e.g., weak network, network disconnection, and specific protocols such as 5G); hardware devices (e.g., specific model sensors, Bluetooth peripherals, and screen resolutions); system versions (e.g., incompatible APIs); third-party dependencies (e.g., SDK version conflicts). 
For example, the Permission module in Android introduced new APIs through APEX updates that functioned correctly on the latest versions (e.g., version 14+) with proper permission enforcement. However, when pushed to Android 13 and earlier systems, where the \codeIn{MANAGE\_DEFAULT\_APPLICATIONS} permission did not exist, these APIs became callable without any permission check, allowing unauthorized apps to modify the user's default browser or assistant settings, producing security issues.

\textbf{Multi-Language/Localization Trigger (Mul/Loc Dep):} Defects triggered by issues related to language, region, or cultural adaptations, including character encoding (e.g., garbled Chinese characters and abnormal emoji displays); time zone/date (e.g., incorrect time calculation caused by daylight saving time transitions); cultural conventions (e.g., validation errors in number formats or address formats). For example, Ubuntu invoked \codeIn{dpkg-L} without explicitly setting the \codeIn{LC\_ALL=C.UTF-8} environment variable but employed strict \codeIn{UTF-8} decoding, causing inconsistent localized output formatting or decoding failures under non-English locales.

\subsubsection{\underline{Severity Level}}
This dimension classifies defects according to the severity of the bugs. The severiy is usually decided by considering multiple aspects, such as the security risk induced by the bug, impact to the usability, the emergency to get a fix, and so on. To achieve an accurate classification of the bug severity typically requires a high expertise about the system implemantaion. As a result, we adopted the severity levels tagged by corresponding developers for each operating system to ensure the reliability of the results. Specifically, Android uses two levels: \textbf{High} and \textbf{Critical}; HarmonyOS uses three levels: \textbf{Low}, \textbf{Medium}, and \textbf{High}; Launchpad uses five levels: \textbf{Undecided}, \textbf{Low}, \textbf{Medium}, \textbf{High}, and \textbf{Critical}. In this study, we analyzed and compared the severity distribution discrepancies across systems without any modifications.

\subsubsection{\underline{Code Level}}

This dimension classifies defects according to the fine-grained code elements that are required to be modified in the given patches for removing the corresponding defects. Different from the other dimensions that one defect only belongs to one category within each dimension, the categories in this dimension are not mutually exclusive since one defect usually involves multiple code changes, such as deleting an \codeIn{Assignment} along with updating an \codeIn{If} statement. In these cases, multiple categories will be assigned to an individual defect. Specifically, in this classification, we adopted the common code elements and structures defined in the programming languages (i.e., C/C++ and Java) used by the studied operating systems. We illustrate all the code elements (i.e., categories) involved in this study as follows, which comprises 12 categories.

\begin{enumerate}
\item \textbf{Entire Class (Class):} Patches involving the \textit{addition} or \textit{deletion} of entire classes or struct definitions (the complete declaration with all members).

\item \textbf{Entire Function (Function):} Patches involving the \textit{addition} or \textit{deletion} of entire function definitions (test function are excluded).

\item \textbf{Function Signature (Signature):} Patches involving the \textit{addition}, \textit{deletion}, or \textit{modification} of function signatures, 
including the function name, parameters (e.g., types), and modifiers (e.g., \codeIn{static}, \codeIn{const}, etc.). The function body is not included. \textbf{Example:} changing \codeIn{void bar(char* s)\{\}} to \codeIn{void bar(const char* s)\{\}}, and  changing \codeIn{int foo(int x)} to \codeIn{int foo(int x, int y)}.

\item \textbf{Loop:} Patches involving the \textit{addition} or \textit{deletion} of entire loop statements (e.g., \codeIn{for} and \codeIn{while}). Modifications to existing loop statments, such as changing the conditions, are not included.

\item \textbf{If:} Patches involving the \textit{addition} or \textit{deletion} of complete \codeIn{if} or \codeIn{else} blocks.
Modifications to the conditional expressions of existing \codeIn{if} statements or partially modifying the if body is not included.

\item \textbf{Exception:}  Patches involving the \textit{addition}, \textit{deletion}, or \textit{modification} of exception handling flows, involving \codeIn{try}, \codeIn{catch}, \codeIn{finally} blocks and \codeIn{throw} statements.
\textbf{Example:} Adding \codeIn{throw new NullPointerException("");} and wrapping code with \codeIn{try\{\}catch(Exception e)\{\}}.

\item \textbf{Function Call (Call):} Patches involving the \textit{addition}, \textit{deletion}, or \textit{modification} of function calls, including changes to the names, arguments, the return value handling, or the call site itself. \textbf{Example:} Adding \codeIn{validate(data);}, replacing \codeIn{malloc(size)} with \codeIn{calloc(1,size)}, and replacing \codeIn{func();} with \codeIn{o=func();}.

\item \textbf{Assignment:} Patches involving the \textit{addition} or \textit{deletion} a complete assignment statement, as well as \textit{modifying} either the left/right hands or operators of existing assignment statements.
\textbf{Example:} 
Adding \codeIn{flag=true;} and changing \codeIn{ptr+=offset;} to \codeIn{ptr+=offset*2;}.

\item \textbf{Return:} Patches involving the \textit{addition}, \textit{deletion}, or \textit{modification} of \codeIn{return} statements, including return value changes and the addition/removal of return points.
\textbf{Example:} Changing \codeIn{return 0;} to \codeIn{return 1;};

\item \textbf{Condition:} Patches involving the \textit{modification} of existing condition expressions only in conditional statements (\codeIn{if}, \codeIn{switch}, etc.) and loop statements (\codeIn{for}, \codeIn{while}, etc.). 
\textbf{Example:} Changing \codeIn{if(a>0)} to \codeIn{if(a>=0)}, \codeIn{while(i>n)} to \codeIn{while(k>n)}, and \codeIn{switch(type)} to \codeIn{switch(type\&MASK)}.

\item \textbf{Object:} Patches involving the \textit{addition}, \textit{deletion}, or \textit{modification} of objects (instances of classes/structs), including their creation (e.g, \codeIn{new} and \codeIn{malloc}), destruction (e.g., \codeIn{delete} and \codeIn{free}), or direct member field modifications. Merely adding or deleting references/pointers to existing objects is not included. \textbf{Example:} 
Adding \codeIn{new Thread(runnable).start();} and removing \codeIn{delete ptr;}.

\item \textbf{Variable:} Patches involving the \textit{addition}, \textit{deletion}, or \textit{modification} of variable declarations (including local/global variables and member fields), involving changes to the variable's type, name, initializer, or modifiers (e.g., \codeIn{final}, \codeIn{volatile}, etc.). Merely updating the access to an existing variable without changing its declaration is not included.
\textbf{Examples:} Changing \codeIn{int MAX=100;} to \codeIn{int MAX=200;} or \codeIn{int a;} to \codeIn{float a;}, and adding \codeIn{int count = 0;}.
\end{enumerate}

As previously noted, the \textbf{Code Level} categories are not mutually exclusive, allowing a single patch to fall into several categories. However, to prevent double-counting, any code elements that reside within newly introduced or removed blocks of compound structures (including Class, Function, Loop, If) are not counted repeatedly.

%% file: analysis.tex
\section{Result Analysis}
\label{sec:result}

\input{tables/all.tex}

\subsection{RQ1: Distribution of OS Defects}
\label{subsec:rq1}

In this section, we analyze the distributions of defects across the three operating systems (i.e., Android, Linux, and HarmonyOS) based on the classification schema introduced in Section~\ref{sec:framework}. The detailed statistical results are presented in Table~\ref{tab:bug_distribution}.
In the following, we will introduce the details in terms of each dimension, and analyze potential reasons.

\subsubsection{OS Layers}
Examining which layer of the operating system contains the most defects reveals clear differences across platforms. In Android, the \textbf{System Service Layer} accounts for the highest share (i.e., 38.80\%), which far exceeds the proportions seen in the other systems. For Linux, the \textbf{Application Layer} contains more than half of all defects, reaching 54.20\%. HarmonyOS differs markedly: the \textbf{Kernel Layer} has the largest share at 71.80\%, and the \textbf{Runtime (or Virtual Machine) Layer} also stands out at 8.80\%, a proportion far higher than in the other two operating systems. These patterns are closely related to each system's design philosophy and its level of maturity. Android's highest defect share in the System Service Layer reflects its mobile-first architecture, which depends heavily on complex services such as permission control and telephony, compounded by frequent updates and vendor customizations. Linux's concentration of defects in the Application Layer arises from its extremely mature and stable kernel, pushing most remaining defects into user-space applications, drivers, and the vast ecosystem of distributable software. In contrast, HarmonyOS, being relatively new and built around a microkernel-like design, shows a dominant defect share in the Kernel Layer because this core component is novel and insufficiently tested at scale. 

\subsubsection{Affected Function} The most affected function also varies significantly by operating systems. In Android, \textbf{Security and Permission Control} accounts for the largest share, reaching about 60.00\%. For Linux, \textbf{Compatibility and Adaptation} holds the highest share at nearly 41.20\%, a category that is not very prominent in the other two systems ($<=$ 3.00\%). In HarmonyOS, the leading categories are \textbf{Memory and Resource Management}, at 22.80\%, and \textbf{Network and Communication}, at 20.00\%; both are substantially higher than in Android or Linux. 
The results indicate that, compared to the Android and Linux systems, defects in HarmonyOS tend to be distributed more evenly across different categories with respect to the affected function. In contrast, defects in Android and Linux are more likely to be dominated by certain types. For example, 60.00\% of affected functions in Android are security and permission related, while the second most affected function (memory and resource management) accounts for only 12.40\%, which is much lower than the top category. A very similar pattern can be observed in Linux. These differences may be largely attributed to the maturity of each system. Specifically, HarmonyOS is a relatively new system compared to the other two, and thus lacks sufficient testing from end users. In contrast, Android and Linux have been deployed for years; in particular, most defects in Linux are related to compatibility issues, while minimal defects affected the memory and resources management.



\begin{figure}[t]
  \centering
  \includegraphics[width=0.9\columnwidth]{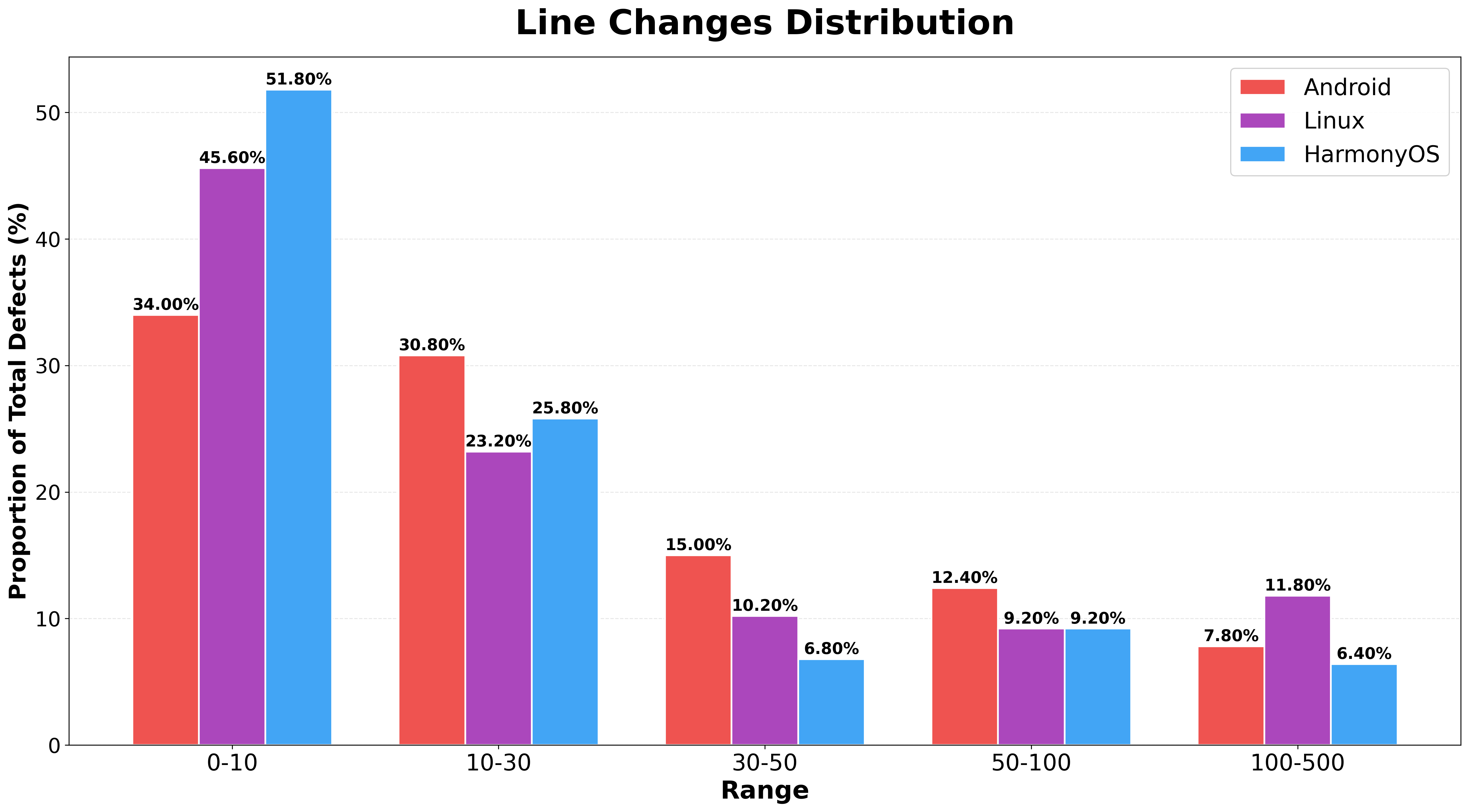}
  \caption{Number distribution of code lines modfied for defect repair.}
  \label{fig:line-changes}
\end{figure}

\subsubsection{Trigger Scenario} The conditions that trigger defects also differ acorss systems. In Android, \textbf{Boundary Condition Trigger} is the highest at 37.00\%, while \textbf{Permission/Identity Dependency Trigger} is also notable at 31.20\%. For Linux, \textbf{Environment Dependency Trigger} dominates at 62.00\%. HarmonyOS again shows a distinct profile: \textbf{Boundary Condition Trigger} is the highest at 52.60\%, and \textbf{Concurrency/Multi‑Thread Trigger} is significantly higher than in the other two systems, reaching 24.00\%. It is worth noting that for Permission/Identity Dependency Trigger, Android’s figure of 31.20\% contrasts sharply with only 1.00\%$\sim$3.00\% in Linux and HarmonyOS. These contrasting profiles reflect fundamental differences in system architecture, execution environments, and design philosophies. Specifically, Android’s defect triggers stem from its sandboxing model and UI-heavy event-driven boundaries, while Linux is dominated by environment dependencies, reflecting its role as a general-purpose OS without a centralized framework. 
HarmonyOS exhibits a distinct hybrid pattern, with the highest boundary condition trigger rate and a notably elevated concurrency and multi‑thread trigger rate, likely arising from its microkernel‑inspired design and newer concurrency model still maturing in comparison to the more established synchronization mechanisms of Android and Linux. However, \textbf{Multi-Language/Localization}  adaptation issues are rare across all systems, thanks to mature internationalization libraries and Unicode adoption. 


\begin{figure}[t]
    \centering
    \includegraphics[width=\columnwidth]{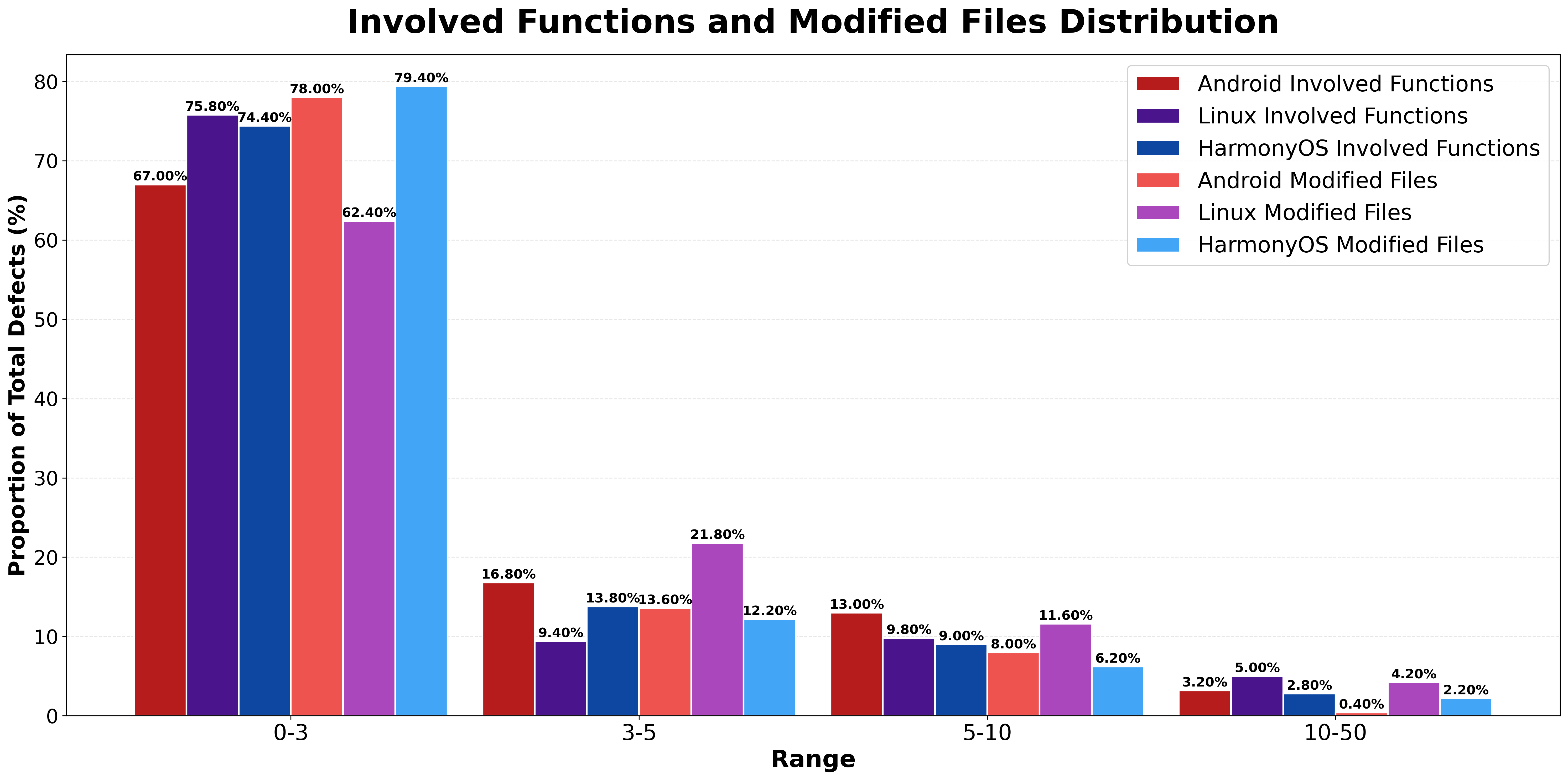}
    \caption{Number distribution of functions and files modfied for defect repair.}
    \label{fig:function-changes}
\end{figure}

\subsubsection{Severity}

In Android and Linux, High-level defects account for the highest proportion, respectively making up 91.00\% and 18.60\% (except for ``Undecided'' for Linux as its large portion of defects do not have associated tags), whereas in HarmonyOS, Medium is the highest, accounting for 51.40\%. We attribute the reasons for this distribution difference multiple reasons. Android’s high rate of \textbf{High}‑severity defects stems from a security‑focused defect tracking system (especially external vulnerability reports), which can also be confirmed by the high propotions of affected functions are related to security and permision.
Linux’s low proportion of \textbf{High}‑severity defects reflects its maturity: most defects are low‑impact (e.g., compatibility issues as explained above). In addition, maintainers reserve ``High'' for true crashes, corruption, or breaches. Particularly, the large share of unclassified defects in Linux stems from a decentralized model—inconsistent triage, users skipping severity ratings, and older tools like Bugzilla that don’t enforce severity selection.
Finally, HarmonyOS’s majority of \textbf{Medium}‑severity rate reflects a deliberate classification philosophy: ``Medium'' is the default for most functional issues (e.g., integration or performance bugs from internal testing). This avoids over‑labeling issues as High or Low, and may broadly cover any defect with a workaround, balancing attention without alarm.

\subsubsection{Repaired Code Elements}
To better understand the bug fixes across systems, we further analyzed the code elements that were repaired to remove them. Specifically, we manually examined all patches for the studied defects and assigned relevant code element tags to each defect, as shown in Table~\ref{tab:bug_distribution}. Across the three systems, the distribution of code elements involved in repairs was similar. The most frequently modified elements were \textbf{if}, \textbf{Function Call}, \textbf{Assignment} and \textbf{Variable}, which aligns with the fix patterns observed in conventional application software~\cite{5328809,10.1145/2610384.2628055}.

Furthermore, to gain a more comprehensive understanding of bug fix complexity, we also analyzed the number of lines modified per repair, as well as the number of functions and files involved. The results are presented in Figures~\ref{fig:line-changes} and~\ref{fig:function-changes}. From these figures, we observe that, regardless of the system, small‑scale fixes affecting no more than ten lines are the most common. This indicates that repairs follow the principle of minimal repair. Additionally, the number of functions and files modified per defect is also concentrated between one and three. This suggests that defect issues are relatively localized, further reflecting the success of modularization in all three systems, an observation consistent with the prevalence of small‑scale repairs.

\begin{finding}
    The distributions of defects
    across the dimensions of OS layers, affected functions, trigger scenarios, and severity 
    differ largely across systems.
    In contrast, the distributions of code elements involved in defect repairs are largely similar across systems.
\end{finding}

\input{tables/correlation}

\subsection{RQ2: Relationship Between Dimensions}
\label{subsec:rq2}

In this section, we introduce the correlations between different classification dimensions within an individual operating system. In particular, we focus on analyzing the relationships among OS layers, affected functions, and trigger scenarios, as these dimensions are essential for developers and maintainers during the debugging process.

\subsubsection{OS Layers vs Affected Functions} To better understand the correlation between OS layers that defects appear and their affected functions, we analyzed the distributions of defects across these two dimensions, the results are presented in Table~\ref{tab:correlation-layer-function}. In the table, for each system, we present the percentage of defects that are belonging to the corresponding categories in terms of the two dimensions. In particular, we omit ``0'' in the table for clarity. 
According to the table, different systems exhibit both commonalities and differences in terms of defect distributions across OS layers and affected functions. On one hand, in most cases, defects occurring at a particular OS layer can affect diverse functions. For example, defects in the System Services layer may lead to compatibility, memory, storage, and other issues. Conversely, the same type of issue can be caused by defects at different layers, such as memory issues can originate from either system libraries or system services. On the other hand, the strength of these correlations also tends to vary across different systems.

In Android, defects are heavily concentrated in \textbf{Security and Permission Control} (Sec/Perm) issues, accounting for 60.00\% of all defects. Among these, about half (28.60\%/60.00\%) occur at the \textbf{System Service} layer, suggesting widespread problems in permission checking and enforcement across system services. The \textbf{Application} and \textbf{Framework} layers also contribute 13.60\% and 12.20\% (out of 60.00\%) of security and permission issues, respectively, reinforcing that permission handling is problematic throughout Android's stack.
In contrast, Linux and HarmonyOS exhibit very different characteristics. In Linux, the majority (9.80\%/14.00\%) of security and permission issues are found at the \textbf{Application} layer, while in HarmonyOS, they are concentrated at the \textbf{Kernel} (8.20\%/18.00\%) layer. 

In Linux, a large portion (41.20\%) of defects lead to \textbf{Compatibility and Adaptability} (Comp/Adp) issues, and the vast majority of these (25.60\% out of the 41.20\%) occur at the \textbf{Application} layer. Furthermore, Linux defects at the \textbf{Framework} layer mainly affect functions related to \textbf{Graphics and Rendering} (Gra/Ren), whereas in the other two systems, the affected functions of framework defects tend to be more varied.

Unlike the other two systems, HarmonyOS exhibits an extreme concentration of defects within the \textbf{Kernel} layer (71.80\%). These kernel defects affect all identified function categories, and their distribution across categories is notably more even. For example, four out of the eight categories each account for over 10.00\% (out of 71.80\%) of kernel-layer defects. In particular, functions related to \textbf{Device Management and Driver} (Dev/Drv) are dominantly affected by kernel-layer defects in HarmonyOS, whereas the causes of similar issues in the other two systems tend to vary.

\begin{finding}
    The correlations between OS layers and affected functions vary widely across operating systems, although some strong correlations exist within individual systems. This suggests that debugging patterns from one system may not always generalize to others.
\end{finding}

\input{tables/correlation2}

\subsubsection{OS Layers vs Trigger Scenarios}

Similarly, this section analyzes the correlations between the OS layers where defects occur and their trigger scenarios. The results are presented in Table~\ref{tab:correlation-layer-trigger}. The findings reflect a similar conclusion: the correlations tend to be diverse across operating systems. For example, most defects (29.80\% out of 54.20\%) from the \textbf{Application} layer were triggered by \textbf{Environment Dependency Issues} (Env Dep) in Linux, whereas the more frequent triggers are \textbf{Permission/Identity Dependency} (Perm/Ide Dep) and \textbf{Boundary Condition} (Bor Con) for Android and HarmonyOS.

This is because Linux is a mature operating system and has undergone extensive testing and real-world deployment, leading to the elimination of many common permission or boundary-related defects at the application layer. What remain are often environment-dependent issues that arise from varying configurations, library versions, or external dependencies, where problems are harder to anticipate in a controlled testing environment. In contrast, Android and HarmonyOS, being more rapidly evolving or relatively newer systems, still exhibit a higher proportion of permission and boundary-condition defects, reflecting ongoing refinement in their permission models and input validation mechanisms.

Likewise, the dominant triggers for kernel defects in Android and HarmonyOS are \textbf{Boundary Condition} (Bor Con) and \textbf{Concurrency/Multi-thread} (Con/Mul Thr) issues, while Linux reveals a completely different distribution, where \textbf{Environment Dependency} (Env Dep) is the most impactful. The reason is also similar. Linux's kernel has matured over decades, with rigorous code review. As a result, many typical boundary condition and concurrency defects have been systematically identified and fixed. In contrast, Android and HarmonyOS, being younger or more rapidly evolving systems, still face ongoing challenges in handling input validation (boundary conditions) and thread safety within their kernel layers.

Nevertheless, defects related to boundary conditions are relatively prevalent across all the three systems and may appear at most OS layers, suggesting that identifying and fixing condition-related defects remains an urgent priority for operating system development.

\begin{finding}
    While correlations between OS layers and defect triggers are system-specific, boundary condition defects are prevalent across all systems and layers, highlighting a universal challenge and underscoring the urgent need to develop targeted defect detection and repair techniques.
\end{finding}

\subsubsection{Affected Functions vs Trigger Scenarios} 

Table~\ref{tab:correlation-function-trigger} presents, for each system, the percentage of defects that overlap across affected functions and trigger scenarios. From the table, we observe a similar phenomenon: \textbf{Boundary Condition} (Bor Con) issues are relatively prevalent, as they can affect all identified functions in our study. This consistency suggests that boundary-related defects are a universal concern across different operating systems.
Nevertheless, the most frequently affected functions by \textbf{Boundary Condition} issues vary across systems. In Android, the most affected functions are security/permission-related (Sec/Perm) and memory/resource-related (Mem/Res Mng) (acounting for 16.40\%/37.00\%, 7.20\%/37.00\%); in Linux, they are security/permission-related(Sec/Perm) and storage-related (Sto Mng) (accounting for 5.80\%/19.00\%, 4.40\%/19.00\%); while in HarmonyOS, they are memory/resouce-related, network/communication-related (Net/Com) and security/permission-related (14.40\%/52.60\%, 10.00\%/52.60\%, 9.80\%/52.60\%). In summary, different systems share both commonalities and discrepancies. Importantly, boundary condition issues are a major root cause affecting the correctness and safety of memory and resource management, since missing boundary checks (e.g., \codeIn{null} pointer checks) are common in practice.

Beyond boundary conditions, other trigger scenarios also exhibit similar cross-system patterns. Specifically, \textbf{Concurrency/Multi-thread}, \textbf{Regular Operation}, and \textbf{State Dependency} triggers tend to affect diverse functions and present comparable phenomena across Android, Linux, and HarmonyOS.

Finally, security and permission-related functions can be affected by a wide range of causes. In addition to well-known reasons such as missing boundary conditions or permission/identity checks, even regular operations may commonly produce security issues. This finding indicates an emerging challenge for ensuring the safety of modern operating systems.

\input{tables/correlation3}

\begin{finding}
    Besides boundary condition defects, concurrency, regular operations, and state dependency triggers also show cross-system patterns. Notably, even regular operations can frequently cause security issues, urging developers to broaden safety checks beyond traditional boundaries.
\end{finding}

\subsection{RQ3: Cross-OS Similarity Analysis}
\label{subsec:rq3}

\begin{figure}[tbp]
\centering
\subfloat[\scriptsize OS Layer]{\includegraphics[width=0.48\linewidth]{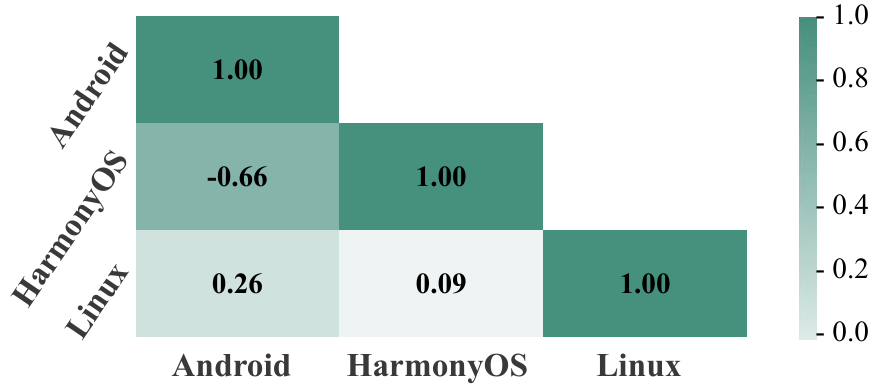}\label{fig:heatmap-layer}}
\hfill
\subfloat[\scriptsize Affected Function]{\includegraphics[width=0.48\linewidth]{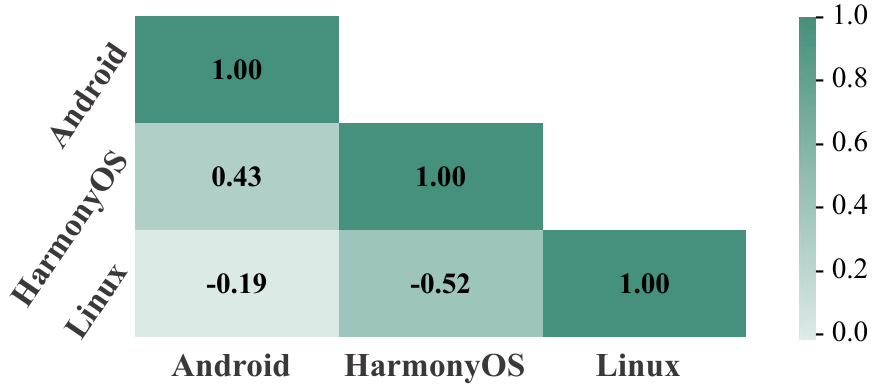}\label{fig:heatmap-function}}

\subfloat[\scriptsize Trigger Scenario]{\includegraphics[width=0.48\linewidth]{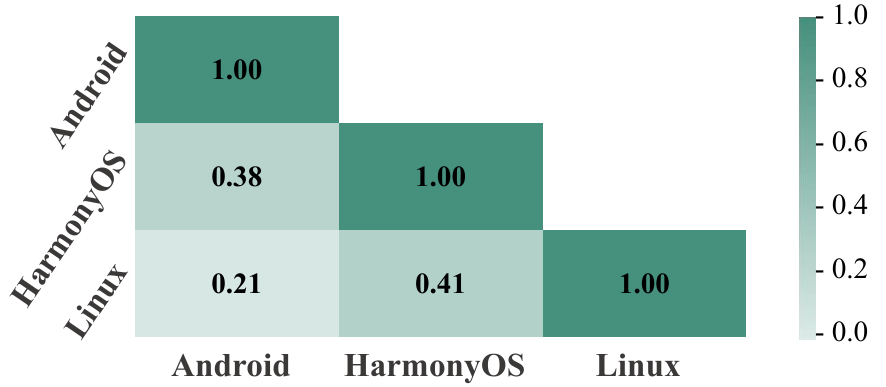}\label{fig:heatmap-trigger}}
\hfill
\subfloat[\scriptsize Overall]{\includegraphics[width=0.48\linewidth]{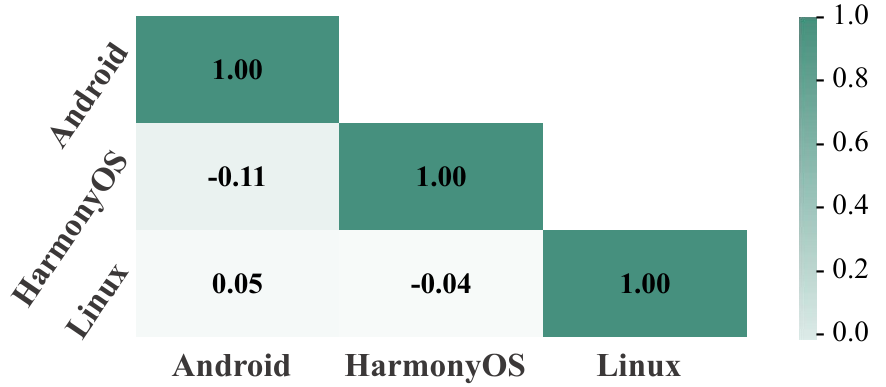}\label{fig:heatmap-all}}

\caption{Similarity heatmaps across systems regarding different dimensions.}
\label{fig:heatmap}
\end{figure}

According to the analysis results, the distributions of defects across the systems tend to vary. To reach a more precise conclusion, this section further quantifies the similarities among these distributions. In particular, we focus on the dimensions of OS layer, affected functions and trigger scenario, while omit severity since different systems adopt varied classification schema. Specifically,
we calculate the Spearman correlation~\cite{Spearman1904} between each pair of OSes to quantify how similar their defect distributions are across different classification dimensions. Spearman’s correlation coefficient is a statistical measure of the strength of a monotonic relationship between two paired variables.
The heatmaps in Figure~\ref{fig:heatmap} visualize these coefficient results, where
[0.8,1.0] indicates the very strong correlation, [0.6,0.8) indicates the strong correlation, [0.4, 0.6)
indicates the moderate correlation, while [0, 0.4) indicates the weak or no correlation. Negative value indicates a negative correlation, with the magnitude interpreted by the same thresholds above.

%


\textbf{Overall Defect Distribution}. As shown in Figure~\ref{fig:heatmap-all}, the three systems exhibit largely different defect distributions overall. The highest absolute coefficient value is merely -0.11 (between Linux and Android), which manifests no substantive monotonic correlation and obvious divergence in defect distribution. This underscores the necessity and urgency of systematically analyzing and comparing OS defects to guide the development of specialized debugging techniques.

\textbf{By OS Layer}. The correlation between HarmonyOS and Android is -0.66, representing strong negative correlation. It indicates that the variation trend of various types of defect counts for HarmonyOS is opposite to Android on the OS layer: when a certain type of defect rises in Android, the quantity of such defects tends to decline in HarmonyOS, and vice versa.
The correlation between Linux and Android is only 0.26, Linux and HarmonyOS reaches 0.09. The absolute values of both coefficients are below 0.4, which means nearly no correlation exists between Linux and the other two operating systems in terms of defect distribution.
Overall, the three operating systems do not share consistent defect distribution patterns at the OS layer. Only Android and HarmonyOS exhibit strong negative correlation, while Linux has almost no correlation with Android and HarmonyOS respectively, demonstrating obvious differentiation in defect composition rules across these operating systems.

\textbf{By Affected Function}. Linux exhibits a moderate negative correlation (-0.52) with HarmonyOS and weak negative correlation (-0.19) with Android regarding affected functions. Linux’s affected functions are mainly concentrated in compatibility and adaptation modules, whose core design goals differ drastically from mobile-oriented Android and distributed HarmonyOS, ultimately resulting in obvious differentiation between Linux and the other two systems. By contrast, Android and HarmonyOS show a mederate positive correlation of 0.43 in terms of affected functions, meaning they possess mild consistency in the types of functions vulnerable to defects. Both systems have substantial defects in Security and Permission Control (Android 60.00\%, HarmonyOS 18.00\%) and Memory and Resource Management (Android 12.40\%, HarmonyOS 22.80\%). This common focus, driven by their shared emphasis on secure, resource-efficient operation in user-facing, always-on environments (Android on mobile devices, HarmonyOS on a distributed, cross-device architecture spanning IoT, smart devices, etc.), explains their moderate correlation.

\textbf{By Trigger Scenario}. Although HarmonyOS and Linux exhibit a moderate positive correlation (0.41)—the strongest correlation among the three system pairs, this value barely reaches the 0.4. This suggests that, despite a limited degree of positive correlation among the three operating systems in the trigger scenario dimension, their overall trigger patterns remain markedly heterogeneous. 

\begin{finding}
    Defect distributions across the three systems differ substantially, indicating the necessity and urgency for better understanding defect features from different systems and developing more targeted debugging techniques.
\end{finding}

%% file: tables/all.tex
\begin{table*}
\centering
\caption{The distributions of the defects regarding different classification dimensions per each OS. In this table, we use shaded colors to highlight the percentage difference, the darker of the color, the larger of the value.}
\label{tab:bug_distribution}
\resizebox{\textwidth}{!}{%
\begin{tabular}{l|cccccccccccc}
\toprule
\multicolumn{13}{c}{\cellcolor{gray!10}\textbf{OS Layer}} \\
\midrule
\textbf{OS} & \textbf{Kernel} & \textbf{SystemLib} & \textbf{RuntimeVM} & \textbf{Framework} & \textbf{SystemService} & \textbf{Application} & & & & & \\ \midrule
\textbf{Android} & \cellcolor{ForestGreen!6}6.00\% & \cellcolor{ForestGreen!20}20.20\% & 0.40\% & \cellcolor{ForestGreen!18}18.00\% & \cellcolor{ForestGreen!39}38.80\% & \cellcolor{ForestGreen!17}16.60\% & & & & & \\
\textbf{Linux} & \cellcolor{ForestGreen!16}15.80\% & \cellcolor{ForestGreen!19}18.80\% & 0.20\% & \cellcolor{ForestGreen!4}3.60\% & \cellcolor{ForestGreen!7}7.40\% & \cellcolor{ForestGreen!54}54.20\% & & & & & \\
\textbf{HarmonyOS} & \cellcolor{ForestGreen!72}71.80\% & \cellcolor{ForestGreen!7}6.80\% & \cellcolor{ForestGreen!9}8.80\% & \cellcolor{ForestGreen!3}2.80\% & \cellcolor{ForestGreen!5}4.80\% & \cellcolor{ForestGreen!5}5.00\% & & & & & \\
\midrule
\midrule
\multicolumn{13}{c}{\cellcolor{gray!10}\textbf{Affected Function}} \\
\midrule
\textbf{OS} & \textbf{Mem/Res Mng} & \textbf{Pro/Tas Sch} & \textbf{Sto Mng} & \textbf{Dev/Drv} & \textbf{Net/Com} & \textbf{Com/Adp} & \textbf{Sec/Perm} & \textbf{Gra/Ren} & & & \\ \midrule
\textbf{Android} & \cellcolor{RoyalBlue!12}12.40\% & \cellcolor{RoyalBlue!5}4.80\% & \cellcolor{RoyalBlue!4}3.60\% & \cellcolor{RoyalBlue!7}7.20\% & \cellcolor{RoyalBlue!3}3.20\% & \cellcolor{RoyalBlue!3}3.00\% & \cellcolor{RoyalBlue!60}60.00\% & \cellcolor{RoyalBlue!6}5.80\% & & & \\
\textbf{Linux} & \cellcolor{RoyalBlue!2}2.00\% & \cellcolor{RoyalBlue!4}3.60\% & \cellcolor{RoyalBlue!8}8.40\% & \cellcolor{RoyalBlue!17}17.20\% & \cellcolor{RoyalBlue!7}6.60\% & \cellcolor{RoyalBlue!41}41.20\% & \cellcolor{RoyalBlue!14}14.00\% & \cellcolor{RoyalBlue!7}7.00\% & & & \\
\textbf{HarmonyOS} & \cellcolor{RoyalBlue!23}22.80\% & \cellcolor{RoyalBlue!6}5.60\% & \cellcolor{RoyalBlue!14}13.80\% & \cellcolor{RoyalBlue!14}13.60\% & \cellcolor{RoyalBlue!20}20.00\% & \cellcolor{RoyalBlue!2}2.40\% & \cellcolor{RoyalBlue!18}18.00\% & \cellcolor{RoyalBlue!4}3.80\% & & & \\
\midrule
\midrule
\multicolumn{13}{c}{\cellcolor{gray!10}\textbf{Trigger Scenario}} \\
\midrule
\textbf{OS} & \textbf{Per/Ide Dep} & \textbf{Reg Ope} & \textbf{Bor Con} & \textbf{Con/Mul Thr} & \textbf{Sta Dep} & \textbf{Env Dep} & \textbf{Mul/Loc Dep} & & & & \\ \midrule
\textbf{Android} & \cellcolor{Goldenrod!31}31.20\% & \cellcolor{Goldenrod!9}8.80\% & \cellcolor{Goldenrod!37}37.00\% & \cellcolor{Goldenrod!7}7.20\% & \cellcolor{Goldenrod!15}14.80\% & \cellcolor{Goldenrod!1}1.00\% & 0.00\% & & & & \\
\textbf{Linux} & \cellcolor{Goldenrod!2}1.60\% & \cellcolor{Goldenrod!9}9.00\% & \cellcolor{Goldenrod!19}19.00\% & \cellcolor{Goldenrod!3}3.00\% & \cellcolor{Goldenrod!5}5.20\% & \cellcolor{Goldenrod!62}62.00\% & 0.20\% & & & & \\
\textbf{HarmonyOS} & \cellcolor{Goldenrod!3}2.80\% & \cellcolor{Goldenrod!7}6.60\% & \cellcolor{Goldenrod!53}52.60\% & \cellcolor{Goldenrod!24}24.00\% & \cellcolor{Goldenrod!11}11.20\% & \cellcolor{Goldenrod!3}2.80\% & 0.00\% & & & & \\
\midrule
\midrule
\multicolumn{13}{c}{\cellcolor{gray!10}\textbf{Severity}} \\
\midrule
\textbf{OS} & \textbf{Critical} & \textbf{High} & \textbf{Medium} & \textbf{Low} & \textbf{Undecided} & & & & & & &\\ \midrule
\textbf{Android} & \cellcolor{Melon!9}8.60\% & \cellcolor{Melon!91}91.00\% & 0.40\% & 0.00\% & 0.00\% & & & & & & &\\
\textbf{Linux} & \cellcolor{Melon!2}2.60\% & \cellcolor{Melon!19}18.60\% & \cellcolor{Melon!11}11.40\% & \cellcolor{Melon!2}2.20\% & \cellcolor{Melon!65}65.20\% & & & & & & &\\
\textbf{HarmonyOS} & 0.20\% & \cellcolor{Melon!22}22.40\% & \cellcolor{Melon!51}51.40\% & \cellcolor{Melon!25}24.80\% & \cellcolor{Melon!1}1.20\% & & & & & & &\\
\midrule
\midrule


\multicolumn{13}{c}{\cellcolor{gray!10}\textbf{Distribution of Repaired Code Elements}} \\
\midrule
\textbf{OS} & \textbf{Class} & \textbf{Function} & \textbf{Signature} & \textbf{Loop} & \textbf{If} & \textbf{Exception} & \textbf{Call} & \textbf{Assignment} & \textbf{Return} & \textbf{Condition} & \textbf{Object} & \textbf{Variable} \\ \midrule
\textbf{Android} & \cellcolor{Violet!3}3.20\% & \cellcolor{Violet!36}35.60\% & \cellcolor{Violet!16}15.60\% & \cellcolor{Violet!4}3.80\% & \cellcolor{Violet!53}53.00\% & \cellcolor{Violet!8}8.20\% & \cellcolor{Violet!63}63.40\% & \cellcolor{Violet!28}28.40\% & \cellcolor{Violet!15}15.40\% & \cellcolor{Violet!19}18.80\% & \cellcolor{Violet!3}2.60\% & \cellcolor{Violet!48}47.60\% \\
\textbf{Linux} & \cellcolor{Violet!2}2.20\% & \cellcolor{Violet!16}15.40\% & \cellcolor{Violet!12}12.20\% & \cellcolor{Violet!4}4.20\% & \cellcolor{Violet!37}37.40\% & \cellcolor{Violet!2}2.20\% & \cellcolor{Violet!50}48.80\% & \cellcolor{Violet!54}54.40\% & \cellcolor{Violet!4}4.40\% & \cellcolor{Violet!24}23.80\% & \cellcolor{Violet!4}3.80\% & \cellcolor{Violet!47}46.80\% \\
\textbf{Harmony} & \cellcolor{Violet!1}1.40\% & \cellcolor{Violet!14}13.60\% & \cellcolor{Violet!8}7.60\% & \cellcolor{Violet!3}3.00\% & \cellcolor{Violet!46}46.00\% & \cellcolor{Violet!0}0.20\% & \cellcolor{Violet!62}61.80\% & \cellcolor{Violet!32}31.80\% & \cellcolor{Violet!20}20.20\% & \cellcolor{Violet!21}21.20\% & \cellcolor{Violet!0}0.40\% & \cellcolor{Violet!38}38.20\% \\
\bottomrule
\end{tabular}
}
\end{table*}

%% file: tables/correlation.tex
\begin{table*}
\centering
\caption{The distributions of defects across OS layers and affected functions. Each cell shows the percentage of defects belonging to the corresponding categories. The larger the value in the cell, the darker of the shaded color. `0' is omitted for clarity.}
\label{tab:correlation-layer-function}
\resizebox{\textwidth}{!}{%
\begin{tabular}{l|cccccccc|cccccccc|cccccccc}
\toprule
 & \multicolumn{8}{c|}{\textbf{Android}} & \multicolumn{8}{c|}{\textbf{Linux}} & \multicolumn{8}{c}{\textbf{HarmonyOS}} \\
 \cline{2-25}
 & \rotatebox{90}{Com/Adp} & \rotatebox{90}{Dev/Drv} & \rotatebox{90}{Gra/Ren} & \rotatebox{90}{Mem/Res Mng} & \rotatebox{90}{Net/Com} & \rotatebox{90}{Pro/Tas Sch} & \rotatebox{90}{Sec/Perm} & \rotatebox{90}{Sto Mng} & \rotatebox{90}{Com/Adp} & \rotatebox{90}{Dev/Drv} & \rotatebox{90}{Gra/Ren} & \rotatebox{90}{Mem/Res Mng} & \rotatebox{90}{Net/Com} & \rotatebox{90}{Pro/Tas Sch} & \rotatebox{90}{Sec/Perm} & \rotatebox{90}{Sto Mng} & \rotatebox{90}{Com/Adp} & \rotatebox{90}{Dev/Drv} & \rotatebox{90}{Gra/Ren} & \rotatebox{90}{Mem/Res Mng} & \rotatebox{90}{Net/Com} & \rotatebox{90}{Pro/Tas Sch} & \rotatebox{90}{Sec/Perm} & \rotatebox{90}{Sto Mng} \\
 \midrule
Application & \cellcolor{ForestGreen!2}0.60 & \cellcolor{ForestGreen!1}0.20 & \cellcolor{ForestGreen!4}1.20 &  &  & \cellcolor{ForestGreen!2}0.60 & \cellcolor{ForestGreen!40}13.60 & \cellcolor{ForestGreen!1}0.40 & \cellcolor{ForestGreen!77}25.60 & \cellcolor{ForestGreen!14}4.60 & \cellcolor{ForestGreen!9}3.00 & \cellcolor{ForestGreen!5}1.60 & \cellcolor{ForestGreen!13}4.40 & \cellcolor{ForestGreen!7}2.20 & \cellcolor{ForestGreen!29}9.80 & \cellcolor{ForestGreen!9}3.00 &  &  & 
\cellcolor{ForestGreen!2}0.80 & \cellcolor{ForestGreen!2}0.60 & \cellcolor{ForestGreen!1}0.20 & \cellcolor{ForestGreen!2}0.80 & \cellcolor{ForestGreen!7}2.20 & \cellcolor{ForestGreen!1}0.40 \\
Framework & \cellcolor{ForestGreen!3}1.00 & \cellcolor{ForestGreen!2}0.60 & \cellcolor{ForestGreen!5}1.80 & \cellcolor{ForestGreen!2}0.80 & \cellcolor{ForestGreen!2}0.60 & \cellcolor{ForestGreen!1}0.40 & \cellcolor{ForestGreen!37}12.20 & \cellcolor{ForestGreen!2}0.60 & \cellcolor{ForestGreen!2}0.80 & \cellcolor{ForestGreen!1}0.20 & \cellcolor{ForestGreen!7}2.40 & &  &  & \cellcolor{ForestGreen!1}0.20 &  & \cellcolor{ForestGreen!1}0.20 &  & \cellcolor{ForestGreen!4}1.20 & \cellcolor{ForestGreen!1}0.40 & \cellcolor{ForestGreen!1}0.40 &  & \cellcolor{ForestGreen!2}0.60 &  \\
Kernel & \cellcolor{ForestGreen!1}0.20 & \cellcolor{ForestGreen!5}1.60 &  & \cellcolor{ForestGreen!6}2.00 & \cellcolor{ForestGreen!2}0.80 &  & \cellcolor{ForestGreen!4}1.40 & & \cellcolor{ForestGreen!16}5.20 & \cellcolor{ForestGreen!20}6.80 & \cellcolor{ForestGreen!3}1.00 &  & \cellcolor{ForestGreen!1}0.40 &  & \cellcolor{ForestGreen!1}0.40 & \cellcolor{ForestGreen!6}2.00 & \cellcolor{ForestGreen!4}1.40 & \cellcolor{ForestGreen!41}13.60 & \cellcolor{ForestGreen!3}1.00 & \cellcolor{ForestGreen!34}11.40 & \cellcolor{ForestGreen!55}18.20 & \cellcolor{ForestGreen!14}4.80 & \cellcolor{ForestGreen!25}8.20 & \cellcolor{ForestGreen!40}13.20 \\
RuntimeVM &  &  &  &  &  &  & \cellcolor{ForestGreen!1}0.40 &  &  &  &  &  &  & \cellcolor{ForestGreen!1}0.20 &  &  & \cellcolor{ForestGreen!1}0.40 &  &  & \cellcolor{ForestGreen!25}8.40 &  &  &  &  \\
SystemLib & \cellcolor{ForestGreen!1}0.40 & \cellcolor{ForestGreen!13}4.20 & \cellcolor{ForestGreen!7}2.20 & \cellcolor{ForestGreen!23}7.60 & \cellcolor{ForestGreen!4}1.20 & \cellcolor{ForestGreen!1}0.20 & \cellcolor{ForestGreen!11}3.80 & \cellcolor{ForestGreen!2}0.60 & \cellcolor{ForestGreen!25}8.20 & \cellcolor{ForestGreen!8}2.80 & \cellcolor{ForestGreen!2}0.60 & \cellcolor{ForestGreen!1}0.20 & \cellcolor{ForestGreen!4}1.20 & \cellcolor{ForestGreen!2}0.80 & \cellcolor{ForestGreen!7}2.40 & \cellcolor{ForestGreen!8}2.60 &  &  & \cellcolor{ForestGreen!2}0.60 & \cellcolor{ForestGreen!5}1.60 & \cellcolor{ForestGreen!1}0.40 &  & \cellcolor{ForestGreen!13}4.20 &  \\
SystemService & \cellcolor{ForestGreen!2}0.80 & \cellcolor{ForestGreen!2}0.60 & \cellcolor{ForestGreen!2}0.60 & \cellcolor{ForestGreen!6}2.00 & \cellcolor{ForestGreen!2}0.60 & \cellcolor{ForestGreen!11}3.60 & \cellcolor{ForestGreen!86}28.60 & \cellcolor{ForestGreen!6}2.00 & \cellcolor{ForestGreen!4}1.40 & \cellcolor{ForestGreen!8}2.80 &  & \cellcolor{ForestGreen!1}0.20 & \cellcolor{ForestGreen!2}0.60 & \cellcolor{ForestGreen!1}0.40 & \cellcolor{ForestGreen!4}1.20 & \cellcolor{ForestGreen!2}0.80 & \cellcolor{ForestGreen!1}0.40 &  & \cellcolor{ForestGreen!1}0.20 & \cellcolor{ForestGreen!1}0.40 & \cellcolor{ForestGreen!2}0.80 &  & \cellcolor{ForestGreen!8}2.80 & \cellcolor{ForestGreen!1}0.20 \\
\bottomrule
\end{tabular}
}
\end{table*}

%% file: tables/correlation2.tex
\begin{table*}
\centering
\caption{The distributions of defects across OS layers and trigger scenarios. Each cell shows the percentage of defects belonging to the corresponding categories. The larger the value in the cell, the darker of the shaded color. `0' is omitted for clarity.}
\label{tab:correlation-layer-trigger}
\resizebox{\textwidth}{!}{%
\begin{tabular}{l|ccccccc|ccccccc|ccccccc}
\toprule

  & \multicolumn{7}{c|}{\textbf{Android}} & \multicolumn{7}{c|}{\textbf{Linux}} & \multicolumn{7}{c}{\textbf{HarmonyOS}} \\ \hline
 & \rotatebox{90}{Bor Con} & \rotatebox{90}{Con/Mul Thr} & \rotatebox{90}{Env Dep} & \rotatebox{90}{Mul/Loc Dep} & \rotatebox{90}{Per/Ide Dep} & \rotatebox{90}{Reg Ope} & \rotatebox{90}{Sta Dep} & \rotatebox{90}{Bor Con} & \rotatebox{90}{Con/Mul Thr} & \rotatebox{90}{Env Dep} & \rotatebox{90}{Mul/Loc Dep} & \rotatebox{90}{Per/Ide Dep} & \rotatebox{90}{Reg Ope} & \rotatebox{90}{Sta Dep} & \rotatebox{90}{Bor Con} & \rotatebox{90}{Con/Mul Thr} & \rotatebox{90}{Env Dep} & \rotatebox{90}{Mul/Loc Dep} & \rotatebox{90}{Per/Ide Dep} & \rotatebox{90}{Reg Ope} & \rotatebox{90}{Sta Dep}\\
 \midrule
Application & \cellcolor{ForestGreen!7}2.40 & \cellcolor{ForestGreen!1}0.40 & \cellcolor{ForestGreen!1}0.40 &  & \cellcolor{ForestGreen!25}8.20 & \cellcolor{ForestGreen!5}1.60 & \cellcolor{ForestGreen!11}3.60 & \cellcolor{ForestGreen!38}12.60 & \cellcolor{ForestGreen!3}1.00 & \cellcolor{ForestGreen!90}29.80 & \cellcolor{ForestGreen!1}0.20 & \cellcolor{ForestGreen!4}1.40 & \cellcolor{ForestGreen!20}6.60 & \cellcolor{ForestGreen!8}2.60 & \cellcolor{ForestGreen!5}1.60 & \cellcolor{ForestGreen!7}1.00 &  &  &  & \cellcolor{ForestGreen!4}1.40 & \cellcolor{ForestGreen!3}1.00\\
Framework & \cellcolor{ForestGreen!16}5.40 &  &  &  & \cellcolor{ForestGreen!20}6.80 & \cellcolor{ForestGreen!8}2.80 & \cellcolor{ForestGreen!9}3.00 &  & \cellcolor{ForestGreen!2}0.60 & \cellcolor{ForestGreen!5}1.80 &  &  & \cellcolor{ForestGreen!1}0.40 & \cellcolor{ForestGreen!2}0.80 & \cellcolor{ForestGreen!3}1.00 & \cellcolor{ForestGreen!3}1.00 &  &  &  & \cellcolor{ForestGreen!2}0.80 & \\
Kernel & \cellcolor{ForestGreen!10}3.20 & \cellcolor{ForestGreen!4}1.40 &  &  &  & \cellcolor{ForestGreen!2}0.60 & \cellcolor{ForestGreen!2}0.80 & \cellcolor{ForestGreen!2}0.60 & \cellcolor{ForestGreen!1}0.40 & \cellcolor{ForestGreen!42}14.00 &  &  & \cellcolor{ForestGreen!2}0.60 & \cellcolor{ForestGreen!1}0.20 & \cellcolor{ForestGreen!100}39.60 & \cellcolor{ForestGreen!58}19.40 & \cellcolor{ForestGreen!8}2.60 &  & \cellcolor{ForestGreen!1}0.40 & \cellcolor{ForestGreen!10}3.40 & \cellcolor{ForestGreen!19}6.40\\
RuntimeVM &  &  &  &  &  & \cellcolor{ForestGreen!1}0.40 &  &  &  &  &  &  & \cellcolor{ForestGreen!1}0.20 &  & \cellcolor{ForestGreen!16}5.40 & \cellcolor{ForestGreen!1}0.40 &  &  &  & \cellcolor{ForestGreen!2}0.80 & \cellcolor{ForestGreen!7}2.20\\
SystemLib & \cellcolor{ForestGreen!40}13.40 & \cellcolor{ForestGreen!12}4.00 &  &  & \cellcolor{ForestGreen!1}0.20 & \cellcolor{ForestGreen!5}1.60 & \cellcolor{ForestGreen!3}1.00 & \cellcolor{ForestGreen!11}3.60& \cellcolor{ForestGreen!2}0.80 & \cellcolor{ForestGreen!38}12.60 &  &  & \cellcolor{ForestGreen!2}0.60 & \cellcolor{ForestGreen!4}1.20 & \cellcolor{ForestGreen!12}4.00 & \cellcolor{ForestGreen!4}1.20 & \cellcolor{ForestGreen!1}0.20 &  & & \cellcolor{ForestGreen!1}0.20 & \cellcolor{ForestGreen!4}1.20\\
SystemService & \cellcolor{ForestGreen!38}12.60 & \cellcolor{ForestGreen!4}1.40 & \cellcolor{ForestGreen!2}0.60 &  & \cellcolor{ForestGreen!48}16.00 & \cellcolor{ForestGreen!5}1.80 & \cellcolor{ForestGreen!19}6.40 & \cellcolor{ForestGreen!7}2.20 & \cellcolor{ForestGreen!1}0.20 & \cellcolor{ForestGreen!11}3.80 &  & \cellcolor{ForestGreen!1}0.20 & \cellcolor{ForestGreen!2}0.60 & \cellcolor{ForestGreen!1}0.40 & \cellcolor{ForestGreen!3}1.00 & \cellcolor{ForestGreen!3}1.00 &  &  & \cellcolor{ForestGreen!7}2.40 &  & \cellcolor{ForestGreen!1}0.40\\

\bottomrule
\end{tabular}
}
\end{table*}

%% file: tables/correlation3.tex
\begin{table*}
\centering
\caption{The distributions of defects across affected functions and trigger scenarios. Each cell shows the percentage of defects belonging to the corresponding categories. The larger the value in the cell, the darker of the shaded color. `0' is omitted.}
\label{tab:correlation-function-trigger}
\resizebox{\textwidth}{!}{%
\begin{tabular}{l|ccccccc|ccccccc|ccccccc}
\toprule

  & \multicolumn{7}{c|}{\textbf{Android}} & \multicolumn{7}{c|}{\textbf{Linux}} & \multicolumn{7}{c}{\textbf{HarmonyOS}} \\ \hline
 & \rotatebox{90}{Bor Con} & \rotatebox{90}{Con/Mul Thr} & \rotatebox{90}{Env Dep} & \rotatebox{90}{Mul/Loc Dep} & \rotatebox{90}{Per/Ide Dep} & \rotatebox{90}{Reg Ope} & \rotatebox{90}{Sta Dep} & \rotatebox{90}{Bor Con} & \rotatebox{90}{Con/Mul Thr} & \rotatebox{90}{Env Dep} & \rotatebox{90}{Mul/Loc Dep} & \rotatebox{90}{Per/Ide Dep} & \rotatebox{90}{Reg Ope} & \rotatebox{90}{Sta Dep} & \rotatebox{90}{Bor Con} & \rotatebox{90}{Con/Mul Thr} & \rotatebox{90}{Env Dep} & \rotatebox{90}{Mul/Loc Dep} & \rotatebox{90}{Per/Ide Dep} & \rotatebox{90}{Reg Ope} & \rotatebox{90}{Sta Dep}\\
 \midrule

Com/Adp & \cellcolor{ForestGreen!4}1.40 &  & \cellcolor{ForestGreen!1}0.20 &  &  & \cellcolor{ForestGreen!1}0.20 & \cellcolor{ForestGreen!4}1.20 & \cellcolor{ForestGreen!5}1.80 & \cellcolor{ForestGreen!1}0.20 & \cellcolor{ForestGreen!107}35.80 & \cellcolor{ForestGreen!1}0.20 &  & \cellcolor{ForestGreen!5}1.60 & \cellcolor{ForestGreen!5}1.60 & \cellcolor{ForestGreen!4}1.40 & \cellcolor{ForestGreen!1}0.20 & \cellcolor{ForestGreen!1}0.40 &  &  & \cellcolor{ForestGreen!1}0.20 & \cellcolor{ForestGreen!1}0.20 \\
Dev/Drv & \cellcolor{ForestGreen!16}5.40 & \cellcolor{ForestGreen!2}0.80 &  &  &  & \cellcolor{ForestGreen!2}0.60 & \cellcolor{ForestGreen!1}0.40 & \cellcolor{ForestGreen!5}1.60 & \cellcolor{ForestGreen!1}0.20 & \cellcolor{ForestGreen!40}13.20 &  &  & \cellcolor{ForestGreen!4}1.60 & \cellcolor{ForestGreen!2}0.60 & \cellcolor{ForestGreen!23}7.60 & \cellcolor{ForestGreen!11}3.60 & \cellcolor{ForestGreen!1}0.40 &  &  & \cellcolor{ForestGreen!1}0.40 & \cellcolor{ForestGreen!5}1.60 \\
Gra/Ren & \cellcolor{ForestGreen!6}2.00 & \cellcolor{ForestGreen!1}0.40 & \cellcolor{ForestGreen!1}0.20 &  &  & \cellcolor{ForestGreen!2}0.80 & \cellcolor{ForestGreen!7}2.40 & \cellcolor{ForestGreen!5}1.60 & \cellcolor{ForestGreen!2}0.80 & \cellcolor{ForestGreen!8}2.80 &  &  & \cellcolor{ForestGreen!3}1.00 & \cellcolor{ForestGreen!2}0.80 & \cellcolor{ForestGreen!3}1.00 & \cellcolor{ForestGreen!6}2.00 & \cellcolor{ForestGreen!1}0.20 &  &  & \cellcolor{ForestGreen!1}0.20 & \cellcolor{ForestGreen!1}0.40 \\
Mem/Res Mng & \cellcolor{ForestGreen!22}7.20 & \cellcolor{ForestGreen!14}4.60 &  &  &  & \cellcolor{ForestGreen!1}0.20 & \cellcolor{ForestGreen!1}0.40 & \cellcolor{ForestGreen!4}1.20 & \cellcolor{ForestGreen!1}0.20 &  &  &  & \cellcolor{ForestGreen!1}0.40 & \cellcolor{ForestGreen!1}0.20 & \cellcolor{ForestGreen!47}14.40 & \cellcolor{ForestGreen!11}3.60 & \cellcolor{ForestGreen!1}0.40 &  &  & \cellcolor{ForestGreen!4}1.20 & \cellcolor{ForestGreen!10}3.20 \\
Net/Com & \cellcolor{ForestGreen!5}1.80 & \cellcolor{ForestGreen!1}0.20 &  &  & \cellcolor{ForestGreen!1}0.40 & \cellcolor{ForestGreen!1}0.40 & \cellcolor{ForestGreen!1}0.40 & \cellcolor{ForestGreen!5}1.80 & \cellcolor{ForestGreen!1}0.20 & \cellcolor{ForestGreen!8}2.60 &  &  & \cellcolor{ForestGreen!2}0.80 & \cellcolor{ForestGreen!4}1.20 & \cellcolor{ForestGreen!30}10.00 & \cellcolor{ForestGreen!19}6.40 & \cellcolor{ForestGreen!2}0.60 &  &  & \cellcolor{ForestGreen!4}1.40 & \cellcolor{ForestGreen!5}1.60 \\
Pro/Tas Sch & \cellcolor{ForestGreen!3}1.00 & \cellcolor{ForestGreen!2}0.80 & \cellcolor{ForestGreen!1}0.20 &  & \cellcolor{ForestGreen!1}0.20 & \cellcolor{ForestGreen!3}1.00 & \cellcolor{ForestGreen!5}1.60 & \cellcolor{ForestGreen!2}0.80 & \cellcolor{ForestGreen!1}0.40 & \cellcolor{ForestGreen!4}1.40 &  &  & \cellcolor{ForestGreen!3}1.00 &  & \cellcolor{ForestGreen!4}1.40 & \cellcolor{ForestGreen!8}2.60 & \cellcolor{ForestGreen!1}0.20 &  &  & \cellcolor{ForestGreen!2}0.60 & \cellcolor{ForestGreen!2}0.80 \\
Sec/Perm & \cellcolor{ForestGreen!49}16.40 & \cellcolor{ForestGreen!1}0.40 & \cellcolor{ForestGreen!1}0.40 &  & \cellcolor{ForestGreen!90}30.00 & \cellcolor{ForestGreen!14}4.80 & \cellcolor{ForestGreen!24}8.00 & \cellcolor{ForestGreen!17}5.80 & \cellcolor{ForestGreen!1}0.20 & \cellcolor{ForestGreen!12}4.00 &  & \cellcolor{ForestGreen!5}1.60 & \cellcolor{ForestGreen!7}2.20 & \cellcolor{ForestGreen!1}0.20 & \cellcolor{ForestGreen!29}9.80 & \cellcolor{ForestGreen!5}1.80 &  &  & \cellcolor{ForestGreen!8}2.80 & \cellcolor{ForestGreen!7}2.20 & \cellcolor{ForestGreen!4}1.40 \\
Sto Mng & \cellcolor{ForestGreen!5}1.80 &  &  &  & \cellcolor{ForestGreen!2}0.60 & \cellcolor{ForestGreen!2}0.80 & \cellcolor{ForestGreen!1}0.40 & \cellcolor{ForestGreen!13}4.40 & \cellcolor{ForestGreen!2}0.80 & \cellcolor{ForestGreen!7}2.20 &  &  & \cellcolor{ForestGreen!1}0.40 & \cellcolor{ForestGreen!2}0.60 & \cellcolor{ForestGreen!21}7.00 & \cellcolor{ForestGreen!11}3.80 & \cellcolor{ForestGreen!2}0.60 &  &  & \cellcolor{ForestGreen!1}0.40 & \cellcolor{ForestGreen!6}2.00 \\

\bottomrule
\end{tabular}
}
\end{table*}

%% file: implication.tex
\section{Implications}
\label{sec:implication}

\subsection{Regarding Testing and Diagnosis}

For Android, defects are highly concentrated in \emph{Security and Permission Control} and the \emph{System Service} Layer, and are mostly triggered by boundary conditions and permission-dependent scenarios. When diagnosing Android defects, developers should first inspect permission management logic, system service components, and edge-case handling paths rather than low-level modules. 
Importantly, it is recommended to use static analysis for permission flows. Tools like PermissionDispatcher\footnote{https://github.com/permissions-dispatcher/PermissionsDispatcher} or custom lint rules can flag missing permission checks before any system service call, and developers should enforce that every permission‑sensitive code path properly handles permission violations. Additionally, fuzzing of system service boundaries is valuable: develop test harnesses that simulate permission toggling, app standby states, and low‑memory conditions while services are active, targeting edge cases such as calling a service one millisecond after permission denial.

For Linux, defects are dominated by \emph{Compatibility and Adaptation} issues in the \emph{Application} Layer, and over 60.00\% are triggered by environment dependencies. Consequently, Linux defect localization should prioritize environment configuration, cross-platform adaptation, and application-layer implementation. Specifically,
environment snapshot and replay should be standard practice. For each application, capture a manifest of dependencies and then use containerization such as Docker to replicate production environments exactly. Tools like Environment Modules\footnote{https://envmodules.io} can help manage multiple library versions. Similarly, automated cross-platform testing is also critical. Developers can use continuous integration pipelines to test against different kernel versions and configuration flags. Finally, developers are encouraged to implement resilience patterns such as graceful degradation when dependencies are missing. For example, falling back to a bundled library or emitting a clear diagnostic message instead of crashing.

For HarmonyOS, more than half of defects appear in \emph{Memory and Resource Management} and the \emph{Kernel} Layer, primarily triggered by boundary conditions and concurrency issues. Troubleshooting should therefore focus on runtime resource scheduling, boundary checks, and multi-thread synchronization mechanisms.
In particular, stress test with HDC (Harmony Device Connector) shell commands to simulate low‑memory scenarios, for instance by triggering memory compaction and launching concurrent applications, is important. Likewise, boundary condition fuzzing is also essential, where developers can use HarmonyOS's distributed testing framework to inject invalid sizes, null pointers, and extreme values into memory allocation APIs. In addition, developers should also employ static analyzers on kernel‑layer modules to catch lock order violations for detecting concurrency issues.




\subsection{Regarding Defect Repair}
Across all three studied systems, the majority of defect-fixing changes are small in scope, requiring fewer than 30 lines of code modification. These changes typically involve limited line additions or deletions, affect only a few files, and touch a small number of functions.
This finding suggests that many defects can be resolved through tightly focused, low-risk edits. Adopting a \textit{minimal-modification} strategy, i.e., making the smallest possible change that correctly addresses the root cause, can reduce the chance of introducing new defects (regressions) compared to larger, more invasive patches. Developers should resist the urge to refactor or clean up unrelated code when fixing a defect, as such collateral changes increase review complexity and fault surface.

Nevertheless, while small, localized changes dominate overall, a significant portion of operating system bug repairs require coordinated modifications that span multiple methods or even multiple files. However, existing automated program repair (APR) techniques are predominantly designed for single-method or single-file changes. This mismatch reveals an urgent need for repair techniques that can handle systemic changes that propagate fixes across function boundaries, shared data structures, or even module interfaces. Most APR benchmarks (e.g., Defects4J~\cite{10.1145/2610384.2628055} and ManyBugs~\cite{7153570}) focus on single-file Java/C++ projects, leaving OS-level repairs largely unaddressed.

%% file: discussion.tex
\section{Threats to Validity}
\label{sec:threats}

The \textit{external threats to validity} primarily stem from the data we used. We systematically collected 500 defects from each of three operating systems, covering the period from April 2023 to December 2025. To enhance the reliability of our conclusions, we selected three distinct operating systems that span both mobile and laptop platforms. These include a well-maintained Linux distribution (representing mature systems) and the emerging HarmonyOS (representing newer systems). To the best of our knowledge, this is the largest-scale study of OS defects to date. Finally, since the sources of bug reports differ across systems, such as Linux bug reports originate from both developers and end‑users, whereas HarmonyOS bug reports are released by official maintainers, the resulting defect distributions may not always reflect the true distribution of defects in real practice.

The \textit{internal threats to validity} 0mainly lie in our manual defect labeling workflow. To mitigate labeling inaccuracies and subjective bias, two authors each with over three years of software development experience independently annotated defects using a generic open-coding framework, consistent with methodologies adopted in numerous prior studies~\cite{Rahman2022,Jiang2019manual,10.1145/3338906.3338955}. In addition, inter-rater agreement was tracked iteratively using Cohen’s Kappa statistic during the whole labeling process. After annotating 100 defect instances (1,500 in total), the Kappa values remained consistently high ($>$0.8), with later iterations reaching 0.9 to 1.0, indicating strong inter-rater consensus. In cases of disagreement between the two primary coders, a third author was consulted for deliberation, and iterative discussions continued until a unified annotation was reached, further enhancing labeling reliability. More importantly, we have offered a clear definition for each category of the classification and open sourced all our studied data for further evaluation and promote future research.





%% file: conclusion.tex
\section{Conclusion}
\label{sec:conclude}

In this paper, we conducted a comprehensive empirical study on defect characteristics across three representative operating systems: Android, Linux, and HarmonyOS. We collected and analyzed 1,500 OS defects from multiple dimensions, including the OS layer where defects occurred, the functions they affected, how they were triggered, their severity, and the code elements involved in their repair. Based on the analysis results, we summarized five key findings. These findings reveal that while some systems share common defect characteristics, the overall distribution of defects differs significantly across the three operating systems, highlighting the necessity and urgency of developing targeted debugging techniques. Finally, based on these findings, we derive a set of implications to facilitate maintenance and guide future research.